\documentclass[aps,prb,twocolumn,showpacs,amsmath,amssymb,superscriptaddress]{revtex4}
\usepackage{graphicx}
\usepackage{dcolumn}
\usepackage{amsmath}
\usepackage{amssymb}
\usepackage{color}
\usepackage{epstopdf}
\newcommand{\eo}[1]{\textcolor{black}{#1}}

\newcommand{\erfc}{\mathrm{erfc}}
\begin{document}
\title{Variational Bethe Ansatz approach for dipolar one-dimensional bosons}
\author{S. De Palo}
\affiliation{CNR-IOM-Democritos National Simulation Centre, UDS Via Bonomea 265, I-34136, Trieste, Italy}
\affiliation{Dipartimento di Fisica Teorica, Universit\`a Trieste, Strada Costiera 11, I-34014 Trieste, Italy}
\author{R. Citro}
\affiliation{Dipartamento di Fisica ``E. R. Caianiello'', Universit\`a
  degli Studi di Salerno, Via Giovanni Paolo II, I-84084 Fisciano (Sa), Italy}
\author{E. Orignac}
\affiliation{Univ Lyon, Ens de Lyon, Univ Claude Bernard, CNRS, Laboratoire de Physique, F-69342 Lyon, France}

\begin{abstract}
We propose a variational approximation to the ground state energy of a
one-dimensional gas of interacting bosons on the continuum based on
the Bethe Ansatz ground state wavefunction of the Lieb-Liniger
model. We apply our variational approximation to a gas of dipolar
bosons in the single mode approximation and
obtain its ground state energy per unit length. This allows for the
calculation of the Tomonaga-Luttinger exponent as a function of
density and the determination of the structure factor at small momenta. Moreover, in the case of attractive dipolar interaction, an
instability is predicted at a critical density, which could be accessed in lanthanide atoms.
\end{abstract}
\date{\today}
\maketitle
\section{Introduction}

One dimensional interacting bosons\cite{cazalilla2011} are a very
 active topic in current research on many-body problem, owing to
 availability of experimental systems and powerful theoretical
 techniques. One-dimensional bosons with repulsive interactions are
 expected to have the Tomonaga-Luttinger liquid
 state\cite{haldane_bosons,giamarchi_book_1d} as the ground
 state, and display simultaneously critical superfluid and density
 wave fluctuations with interaction dependent exponents. In contrast
 to the case of fermions\cite{giamarchi_book_1d}, these exponents cannot be obtained from
 perturbation theory in the vicinity of the non-interacting
 point. Instead, it is necessary to know the dependence of the
 ground state energy per unit length as a function of particle
 density.\cite{haldane_bosons} In the case of integrable
 models\cite{lieb_bosons_1D,amico04_boson_integrable_review} such
 dependence can be obtained analytically, but in the general case,
 one resorts to numerical methods such as Quantum Monte
 Carlo\cite{mazzanti07_1d_hard_rods,citro06_dipolar1d} or Density
 Matrix Renormalization
 Group.\cite{kuhner_bose_hubbard_critical_point} Variational methods
 have also been proposed\cite{hellberg_tj,kawakami_comment_1992,hellberg_hellberg_1992,fradkin1993,pham_laughlin}, using as variational
 wavefunction the ground state wavefunction of the Calogero-Sutherland
 model.\cite{calogero69_model1,calogero69_model2,sutherland71_model1,sutherland71_model2,sutherland71_model3} With
 such variational wavefunctions, the Tomonaga-Luttinger exponent is
 the variational parameter. This form of variational wavefunction can
 be interpreted\cite{capello2007} in terms of a
 Jastrow\cite{jastrow_fonctions} factor.
\eo{ Because of the difficulty of computing
 correlation functions, the use of Bethe Ansatz wavefunctions  
 as variational wavefunctions has been mainly restricted  to few body
 systems\cite{rubeni2012,wilson2014} in harmonic traps, although a Bethe Ansatz Density
 Functional Theory has been proposed in the case of spin-1/2 fermions
 in harmonic potential.\cite{xianlong05_ba_lda,schenk2008} However,
 \cite{caux05_heisenberg_correlations1d,caux_density}
 using determinant representations of correlation functions has allowed
 calculation of the structure factor of Bethe Ansatz integrable
 models.} Such developments enable
 the use of Bethe Ansatz wavefunctions in a variational
 approach.\cite{claeys2017} Moreover, in the case of the
 integrable Lieb-Liniger\cite{lieb_bosons_1D} gas, an
 approximation to the exact structure factor\cite{caux_density} is
 known, that further simplifies the variational calculation \eo{in the
 thermodynamic limit}.\\
 Here we introduce a variational approach using the Bethe-Ansatz wavefunctions of the Lieb-Liniger model as variational wavefunctions to determine the Tomonaga-Luttinger exponents of a one-dimensional interacting model of bosons with a sufficiently short-range interaction. In particular, we apply it to a dipolar gas, using the separation of the dipole-dipole interaction(DDI) in an effective contact potential and a soft long-range part. \\
 This study is particular timely as, recently, highly magnetic lanthanide atoms such as Dysprosium
and Erbium have given access to strong magnetic dipole-dipole
interactions (DDI) in ultracold
atomic physics .\cite{Lev_2011,Lev_2012,Ferlaino_2012,Ferlaino_2014} The interplay of the short-ranged
Van Der Waals s-wave interaction and the long-range and
anisotropy nature of DDI in the atomic gas has enabled the
exploration of a wide variety of phenomena. The most recent ones are  novel quantum
liquids,\cite{Pfau_2016,Ferlaino_2016,Pfau_prl_2016} strongly correlated lattice states,\cite{Tolra_2013,Tolra_2016,Ferlaino_science_2016}
exotic spin dynamics\cite{tolra_spin_2016} and the emergence of
thermalization in a nearly integrable quantum gas.\cite{tang2018}
An even more exciting physics can be accessed when dimensionality is reduced. In fact in optical lattices, one would
be able to create dipolar Tomonaga-Luttinger liquids\cite{santos_cond_07,citro06_dipolar1d}
as well as novel quantum phases,\cite{Sun_2007} including analogs
of fractional quantum Hall states.\cite{Hafezi_2007} On the application side, setting the DDI strength to zero
 improves the sensitivity of atom interferometry,\cite{Fattori_2008},
 \eo{atomtronic devices based on dipolar interactions have been
 proposed\cite{wilsmann2018},}   and tuning the \eo{interaction} strength
from positive to negative may 
find application in the simulation of nuclear matter.

Special attention has been devoted to the strictly one-dimensional
case with repulsive interactions, in
which both the determination of the equation of state~\cite{Lozovik} and of the structure factor\cite{citro07_luttinger_hydrodynamics}
has suggested the existence
of a crossover from a liquid-like, superfluid state with the
characteristics of a Tonks-Girardeau gas~\cite{TonksG} to a
quasi-ordered (particles localized at lattice sites), normal-fluid
state with increasing the linear density. However, in trapped atom
experiments, the finite transverse width of the trap allows
an averaging of
repulsive and attractive dipolar
interactions\cite{santos_cond_07,Reimann} that precludes the
formation of the quasi-ordered state.
In the present paper, we wish to
understand the crossover from the low density Tonks-Girardeau-like
regime to the quasi-BEC regime at high density, by studying the
evolution of the Tomonaga-Luttinger exponent with an approach
applicable to interacting one-dimensional boson models in the
continuum. Understanding the interplay between short-range  Van Der
Waals and longer ranged  interactions  remains an  experimental
challenge that may lead to new physics.

The paper is organized as follows. Section
\ref{sec:variational_approach} describes the variational approach for
a general setting. Section~\ref{sec:themodel} introduces a model of
 dipolar bosons in a transverse harmonic
 trap,\cite{santos_cond_07,Reimann} and discusses the application of
 the variational approach and the computation of
the Tomonaga-Luttinger parameters. In Section~\ref{sec:conclusions} we offer
some conclusions  and perspectives.

\section{The variational approach}
\label{sec:variational_approach}
We consider the full Hamiltonian of a one-dimensional interacting
bosonic system
\begin{equation}\label{hamiltonian}
 H_{1D}=\mathcal{K}+\mathcal{V},
\end{equation}
where
\begin{eqnarray*}
\mathcal{K}=-\frac{\hbar^2}{2m}\sum_{i=1}^{N}\frac{\partial^2}{\partial x_j^2}
, \hskip0.51cm
\mathcal{V}=\!\!\sum_{1\le i < j \le N} v(x_i-x_j)
\end{eqnarray*}
with $v(x)$ a sufficiently short ranged interaction, with
Fourier transform
$  \hat{v}(k)=\int dx v(x) e^{-ik x} $,
defined for all $k$.

We then introduce the variational Hamiltonian:
\begin{equation}\label{var-ham}
H_{\mathrm{var}}=\mathcal{K}+g \mathcal{U},
\end{equation}
where
\begin{equation}
\nonumber
  \mathcal{U}=\sum_{1\le i < j \le N} \delta(x_i-x_j)
\end{equation}
which is a Lieb-Liniger Hamiltonian that is Bethe-Ansatz integrable\cite{lieb_bosons_1D,lieb_excit}.
Let $|\psi_0(g)\rangle$ the ground state wavefunction of the Lieb-Liniger Hamiltonian~(\ref{var-ham}), such that
$H_{\mathrm{var}}|\psi_0(g)\rangle=E_0 (g)|\psi_0(g)\rangle$.

We re-write the original Hamiltonian (\ref{hamiltonian}) in terms of the  variational one as
\begin{equation}
H_{1D}=H_{var}+\mathcal{V}-g \mathcal{U}
\label{ham_V_U}
\end{equation}
and use $|\psi_0(g)\rangle$ as a variational wavefunction, so that
$E_{var}=\langle\psi_0(g)|H_{1D}|\psi_0(g)\rangle$ is the variational
energy\cite{claeys2017} to be minimized as a function of $g$
\begin{equation}
\nonumber
E_{var}(g)=E_0(g)+\langle\psi_0(g)|\mathcal{V}-g \mathcal{U}|\psi_0(g)\rangle.
\end{equation}
The Hellmann-Feynman theorem,
$\frac{\partial E_0(g)}{\partial
  g}=\langle\psi_0(g)|\mathcal{U}|\psi_0(g)\rangle$, yields
\begin{equation}\label{var-energy2}
E_{var}(g)=E_0(g)-g \frac{\partial E_0(g)}{\partial g}+\langle\psi_0(g)|\mathcal{V}|\psi_0(g)\rangle.
\end{equation}
We then express  $\langle\psi_0(g)|\mathcal{V}|\psi_0(g)\rangle$ in terms
of  the static structure factor $S(k)$ (see App.~\ref{app:pcf}
for derivation and definitions) in the ground state
$|\psi_0(g)\rangle$ and obtain the trial energy per unit length $e_{var}(g)=E_{var}(g)/L$:
\begin{eqnarray}\label{var-energy3}
\nonumber
&&e_{var}(g)=e_0(g)-g \frac{\partial e_0(g)}{\partial g}+\frac n 2 \lbrack n \hat{v}(k=0)\\
&&+ \int_{-\infty}^{\infty} \frac{dk}{2\pi} \hat{v}(k) S(k)-v(x=0) \rbrack,
\end{eqnarray}
where $n=N/L$ is the number of bosons per unit length and
$e_0(g)=E_0(g)/L$ is the Lieb-Liniger energy per unit length.

The calculation of the variational energy requires knowledge of 
$e_0(g)$ and the static structure factor $S(k)$ of the Lieb-Liniger model.
The ground-state energy
can be obtained from the Bethe Ansatz
solution\cite{lieb_bosons_1D} by solving an integral equation (see App.~\ref{app:Lieb_liniger_results}). It
takes the form
\begin{equation}
  \label{eq:e0-lieb-liniger}
  e_0(g)=\frac{\hbar^2 n^3}{2 m} \epsilon_{LL}(\gamma),
\end{equation}
with the dimensionless parameter
\begin{equation}
  \label{eq:gamma-def}
  \gamma=\frac{m g}{\hbar^2 n}.
\end{equation}
\eo{Moreover, an analytical
conjecture for the exact result for $\epsilon_{LL}$ is known,\cite{lang2017,ristivojevic2019,marino2019}
(see App.~\ref{app:Lieb_liniger_results})
and can be  readily used
to estimate the derivative in Eq.~(\ref{var-energy3}).} 
The static structure factor $S(k)$ has been obtained from the form factor expansion\cite{caux_density}, or using Monte Carlo approaches\cite{Astrakharchik} for
selected interaction strengths. Later, an analytic expression of
$S(k)$, interpolating between weak and strong repulsion and  in broad agreement
with the result of Ref.~\onlinecite{caux_density} was
proposed.\cite{cherny2008} Using that approximation greatly simplifies
the evaluation of the trial energy in Eq.~(\ref{var-energy3}).
Once the optimal $g$ that minimizes the trial energy has been found,
Eq.~(\ref{var-energy3}) yields an approximation for the ground state
energy per unit length $e_{GS}$
of the original Hamiltonian.

A one dimensional interacting system of spinless bosons with repulsive interactions is expected to
form a Tomonaga-Luttinger liquid in its ground
state.\cite{cazalilla2011} The low-energy excitations of such
Tomonaga-Luttinger liquid are
described by a  bosonized Hamiltonian
\begin{equation}
  \label{eq:bosonized}
  H_b=\hbar \int \frac{dx}{2\pi} \left[ u K (\pi \Pi)^2 + \frac u K
    (\partial_x \phi)^2\right],
\end{equation}
where $u$ is the velocity of excitations and $K$ the Tomonaga-Luttinger exponent.
\cite{haldane_bosons}
\eo{The latter exponent determines the long distance decay of the single
particle Green's
function and of the density-density correlations\cite{haldane_bosons}
as well as the stability against the formation of a gapped state when
an infinitesimal periodic potential commensurate with the density
is applied along
the tubes.\cite{haller2010,dalmonte2010,boeris2016} The effect of
renormalization of the Tomonaga-Luttinger exponent by a finite
periodic potential on the phase diagram has been studied in
Ref.~\onlinecite{boeris2016}.}    
As a result of Galilean invariance\cite{haldane_bosons}
of (\ref{hamiltonian}),
\begin{equation}
  \label{eq:galilean}
   u K =\frac{\hbar \pi n}{m},
\end{equation}
while\cite{haldane_bosons}
\begin{equation}\label{eq:compress}
\frac u K = \frac 1 {\pi\hbar}  \frac{\partial^2 e_{GS}}{\partial n^2},
\end{equation}
where $e_{GS}(n)$ is the ground state energy per unit length. The
existence of the Tomonaga-Luttinger liquid requires $\frac{\partial^2
  e_{GS}}{\partial n^2}>0$. 
The vanishing of $\frac{\partial^2 e_{GS}}{\partial n^2}$  signals a collapse  instability\cite{nakamura1997,cabra_instabilityLL} towards a state of high (possibly infinite) density.

In the next section, we will illustrate the application of this
variational method to a gas of one-dimensional dipolar bosons with
repulsive contact interaction.

\section{The quasi-one dimensional dipolar model}
\label{sec:themodel}
The effective Hamiltonian in the single-mode approximation
(SMA)\cite{santos_cond_07,Reimann}, for polarized bosonic dipoles trapped in
quasi-one dimensional geometry reads:
\begin{eqnarray}
\nonumber
H_{Q1D}&=&-\frac{\hbar^2}{2m} \sum_i \frac{\partial^2}{\partial x_i^2} +
\sum_{i<j} V_{Q1D}(x_i-x_j) \\
&& + g_{VdW}\sum_{i<j}\delta (x_i-x_j)
\label{eq:Ham_SMA}
\end{eqnarray}
where, compared to Hamiltonian~(\ref{hamiltonian}) we have two
  contributions to the two-body potential energy: $V_{Q1D}(x)$ is the
  effective 1D dipole-dipole interaction obtained after projection of the
transverse degrees of freedom\cite{santos_cond_07,Reimann} and  $g_{VdW}\delta(x)$, which originates from the Van der Waals interaction and other
short range interatomic interactions. These latter interactions are represented in three dimensions by the Huang-Yang pseudopotential.\cite{huang1957,olshanii_cir}
After projection of the transverse degrees of freedom, the pseudopotential
yields the one-dimensional contact interaction with\cite{olshanii_cir}
 $g_{VdW}=\frac{2\hbar^2 a_{3D}}{m l_\perp^2}
 \left(1-\mathcal{C}\frac{a_{3D}}{\sqrt{2} l_\perp}\right)^{-1}$
where $a_{3D}$ is the s-wave scattering length of the
three-dimensional short range potential,
$l_\perp=\sqrt{\hbar/(m \omega_\perp)}$ is the transverse trapping
length, and $\mathcal{C}=1.4603\ldots$ is a numerical constant. Away
from the confinement induced resonance, $a_{3D} \ll l_\perp$, the
single mode approximation is applicable, and one can approximate
\begin{equation}\label{eq:contactSMA}
  g_{VdW}=\frac{2\hbar^2 a_{3D}}{m l_\perp^2}.
\end{equation}
If one wished to include confinement induced resonances, it would be
necessary to include the Van der Waals and the dipolar interaction in
a single Huang-Yang pseudopotential.\cite{shi2014} Such treatment is
beyond the scope of our manuscript.

In the Hamiltonian Eq.~(\ref{eq:Ham_SMA}), the effective 1D dipole-dipole
interaction is\cite{santos_cond_07,Reimann}
\begin{eqnarray}
&&V_{Q1D}(x)=V(\theta)
\left[ V^{1D}_{DDI} \left(\frac{x}{l_\perp}\right)
-\frac{8}{3}\delta \left(\frac{x}{l_\perp}\right) \right], \\
&&V(\theta)=\frac{\mu_0\mu^2_D}{4 \pi}\frac{1-3 \cos^2{\theta}}{4 l^3_\perp}, \\
&&V^{1D}_{DDI} \left(\frac{x}{l_\perp}\right)=-2\left| \frac{x}{l_\perp}\right|
+\sqrt{2\pi} \left[ 1+ \left(\frac{x}{l_\perp}\right)^2 \right]
\nonumber
\\
&&\times e^{\frac 1 2 \left(\frac{x}{l_\perp}\right)^2} \erfc \left[\left|
   \frac{x}{\sqrt{2} l_\perp}\right| \right],
\label{V1d_dd1}
\end{eqnarray}
where $\mu_0$ is the magnetic permeability of the vacuum, $\mu_D$ is
the magnetic dipolar moment of the atom ($\mu_D=9.93\mu_B$ in the case\cite{tang2018} of
${}^{162}$Dy) and $\theta$ is the angle of the dipoles with respect to the $x$-axis.
The Fourier transform of the soft-dipolar interaction
$V_{DDI}^{1D}(r/l_\perp)$ reads\cite{santos_cond_07}:
\begin{equation}\label{Vddi-FT}
\hat{V}_{DDI}(k)=4 l_\perp \left[ 1-\frac{(k
  l_\perp)^2}{2}e^{\frac{(k l_\perp)^2}{2}}E_1\left( \frac{(k
    l_\perp)^2}{2}\right)\right],
\end{equation}
where $E_1(x)=\Gamma(0,x)$ is the exponential integral
function\cite{abramowitz_math_functions}.

We can divide the Hamiltonian, Eq.~(\ref{eq:Ham_SMA}), into a Lieb-Liniger part that contains all the
contact interactions and the rest, a non-integrable soft-dipolar part.
The Lieb-Liniger part of the Hamiltonian then reads:
\begin{eqnarray}
&&H^{LL}_{Q1D}=-\frac{\hbar^2}{2m}\sum_i \frac{\partial^2}{\partial x_i^2}
+g_{Q1D}(\theta)\sum_{i<j} \delta(x_i-x_j)
\nonumber
\end{eqnarray}
where $g_{Q1D}(\theta)=
\left[ g_{VdW}-V(\theta)\frac{8}{3} l_\perp \right]$. The
dimensionless parameter $\gamma_0$ associated to $g_{Q1D}(\theta)$ is
given by:
\begin{eqnarray}
\label{eq:eff_gamma}
\nonumber
\gamma_0&=&\frac{1}{n}\frac{m}{\hbar^2}g_{Q1D}(\theta)=\frac{2}{n a_{Q1D}}=
\gamma_{vdW}+\gamma_{d}=\\
&=& \frac{2}{n}\left[ -\frac{1}{a_{1D}}- 4\frac{a_d}{l^2_\perp}
\frac{1-3\cos^2{\theta}}{3}\right]
\end{eqnarray}
where we have introduced
the dipole length scale $a_d$ for
the one dimensional dipolar interaction, as defined in Ref.~\onlinecite{tang2015}, which
for the ${}^{162}Dy$ atoms is $
a_d=\frac{\mu_0 \mu^2_D m}{8\pi \hbar^2}\simeq 195 a_0 $ where $a_0$
is the Bohr radius. In the following, we will treat $a_{1D}$ as a
phenomenological parameter measuring the strength of the short range
interaction. This value can be converted into a three-dimensional
interaction using the approximation of Eq.~(\ref{eq:contactSMA}).

The original Hamiltonian, Eq.~(\ref{eq:Ham_SMA}), can therefore be written as the sum of the Lieb-Liniger
part and the rest:
\begin{eqnarray}\label{eq:decomposition}
H=H^{LL}_{Q1D}(\gamma_0)+ V(\theta) \sum_{i<j}V^{1D}_{DDI}\left(\frac{x_i-x_j}{l_\perp}\right).
\end{eqnarray}
We can now use a variational Hamiltonian of the form Eq.~(\ref{var-ham}), in which the contact interaction contains already the short range part of~(\ref{eq:Ham_SMA}),
and minimize the trial ground state energy~(\ref{var-energy3}). The variational contact interaction is
written $g=g_{Q1D}(\theta)+\bar{g}$, with $\bar{g}$ to be determined by minimization.

Following Sec.~\ref{sec:variational_approach}, we can extract the optimal correction to $g_0=g_{Q1D}(\theta)$
by minimizing the trial energy per particle  written in terms of
dimensionless ratios $a_d/l_{\perp},n l_{\perp}$ and the adimensional
 parameter $\gamma=\frac{1}{n}\frac{m}{\hbar^2}\left[ g_{Q1D}(\theta)+
\bar{g}\right]=\gamma_0+\bar{\gamma}$ with respect to $\bar{\gamma}$
\begin{eqnarray}
\label{eq:evar_k}
&\!\!&\frac{2m E}{N\hbar^2 n^2}=\epsilon(\gamma)-\bar{\gamma}\frac{\partial \epsilon(\gamma)}{\partial \gamma}
+ 2 \frac{a_d}{l_\perp} \frac{1-3 \cos^2{\theta}}{n l_\perp}\times \\
&&
 \left\{ 1 +\int^\infty_0 \!\! dq \left[ S(\pi q n;\gamma)-1 \right]
   \right. \times
   \nonumber \\
&& \left. \left[1- \frac{\pi^2 q^2 n^2 l^2_\perp}{2} e^{\frac{\pi^2 q^2 n^2 l^2_\perp}{2} }
\Gamma \left( 0,\frac{\pi^2 q^2 n^2 l^2_\perp}{2} \right) \right] \right\},
\nonumber
\end{eqnarray}
where $q$ is dimensionless.
Using for instance the golden search algorithm,\cite{press92_golden}
or simply scanning the energy as
a function of $\bar{\gamma}$ it is possible to find the minimum of
Eq.~(\ref{eq:evar_k}).
The optimal $\gamma$, obtained through the minimization procedure,
depends on three independent dimensionless parameters, such as $n
a_{1D}$, $a_{1D}/l_\perp$ and $a_d(1-3\cos^2\theta)/l_\perp$. At the critical angle
\begin{equation}\label{critangle}
  \theta_c =\arccos\left(\frac 1 {\sqrt{3}}\right)
\end{equation}
the optimal $\gamma$ coincides by construction
with  $\gamma_0$ and depends only on $n a_{1D}$.
At large densities
(see App.~\ref{app:high_and_low}) the contribution of the dipolar interactions in Eq.~\ref{eq:evar_k} is depressed by the
$1/(n l_\perp)$ factor together with the decay of $\hat{V}(q \pi nl_\perp) \sim 2/(q \pi n l_\perp)^2$,
so that upon minimization $\gamma \rightarrow \gamma_0$
asymptotically, as if the dipolar interaction was reduced to its
contact contribution.
If we consider Eq.~(\ref{eq:eff_gamma})  the contribution of the contact
term  $\gamma_d$ due to dipolar interactions is independent of density
$n$, positive when
$\theta<\theta_c$ and negative for $\theta>\theta_c$, up to a point
where $\gamma_0$ can change its sign and become negative. In such a
case, the Tomonaga-Luttinger liquid state should be unstable at high
density.
By contrast, for low densities, the non-contact contribution of the
dipolar interactions to the variational energy is enhanced, and
overwhelms  the contact term. The optimal $\gamma$ is enhanced
when $\theta>\theta_c$ and lowered when $\theta<\theta_c$ (see
App.~\ref{app:minimization}).

\begin{figure}[h]
\begin{center}
\includegraphics[height=65.mm]{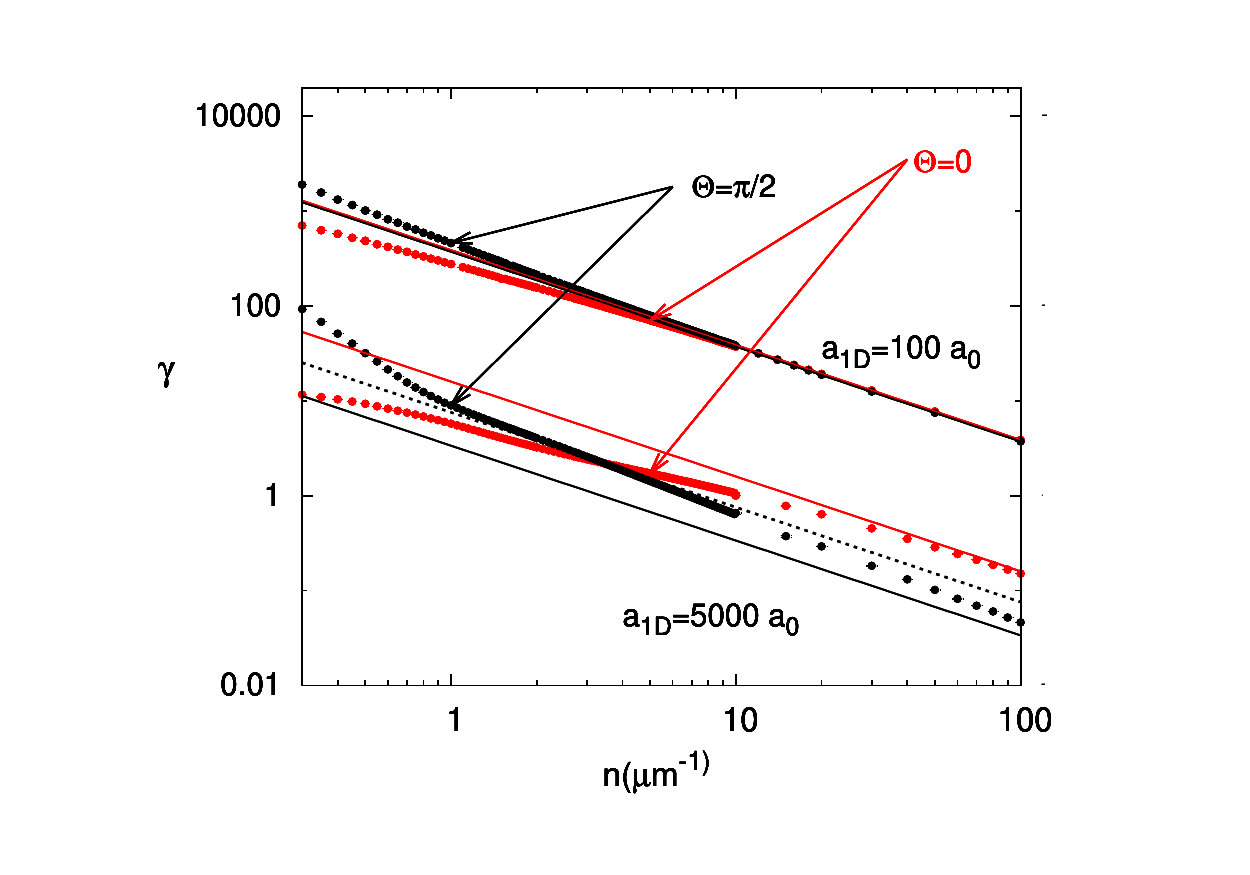}
\end{center}
\caption{(Color online) $\gamma$ as a function of $n$ for two values of the
  scattering length for contact interaction $a_{1D}$. Solid lines
  represent $\gamma_0$ from Eq.~(\ref{eq:eff_gamma}) at angles
  $\theta=0$ (red) and $\theta=\frac \pi 2$ (black). The dashed (black)
  line represents $\gamma_{vdW}$. The diamonds represent the
  optimal $\gamma$ obtained by minimizing Eq.~(\ref{eq:evar_k}), for angle $\theta=0$ (red) and
  $\theta=\frac \pi 2$ (black). At large density, $\gamma$ admits  $\gamma_0$ as asymptote. }
 \label{fig:gamma_n_a1d}
 \end{figure}

In Fig.~\ref{fig:gamma_n_a1d} we show the dependence of $\gamma$ (solid dots) on the density $n$ for
two selected cases, $a_{1D}=100 a_0$ and $a_{1D}=5000 a_0$, and two angles, $\theta=0$ and $\theta=\pi/2$.
For comparison the value of $\gamma_0$ (solid lines) from Eq.~(\ref{eq:eff_gamma}) are shown, as well
as $\gamma_{vdW}=2/(n a_{1D})$ (dashed lines).
When not stated explicitly all results refer to estimates where we have taken
$a_d=195 a_0$ and $l_\perp=57.3 nm$.\cite{tang2018}

\begin{figure}[h]
\begin{center}
\includegraphics[height=65.mm]{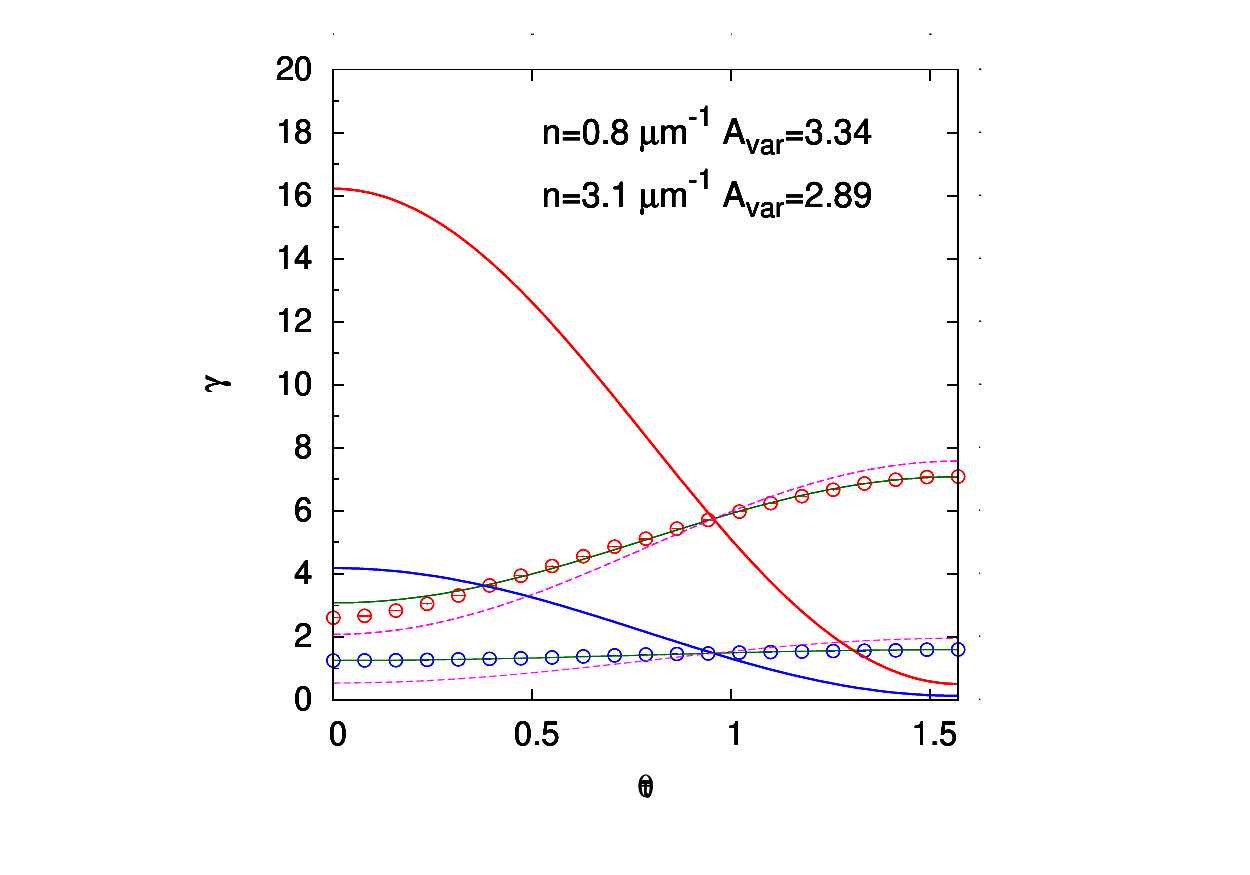}
\end{center}
\caption{$\gamma$ as a function of $\theta$ for $n=0.8 \mu m^{-1}$ red curves,
and for $n=3.1 \mu m^{-1}$, blue curves. Solid curves represent $\gamma_0$ while open dots
are the results of the minimization procedure. The dashed magenta
lines are the estimates
for $\gamma$ if we use Eq.~(\ref{eq:geff-def}) with $A=3.6$ for both
densities\cite{tang2018}, while the dark-green lines fits of the
variational estimates to an expression of the form~(\ref{eq:geff-def})
with $n$ dependent values of $A$  indicated in the inset.}
\label{fig:gamma_var}
\end{figure}

\begin{figure}
  \centering
  \includegraphics[width=9cm]{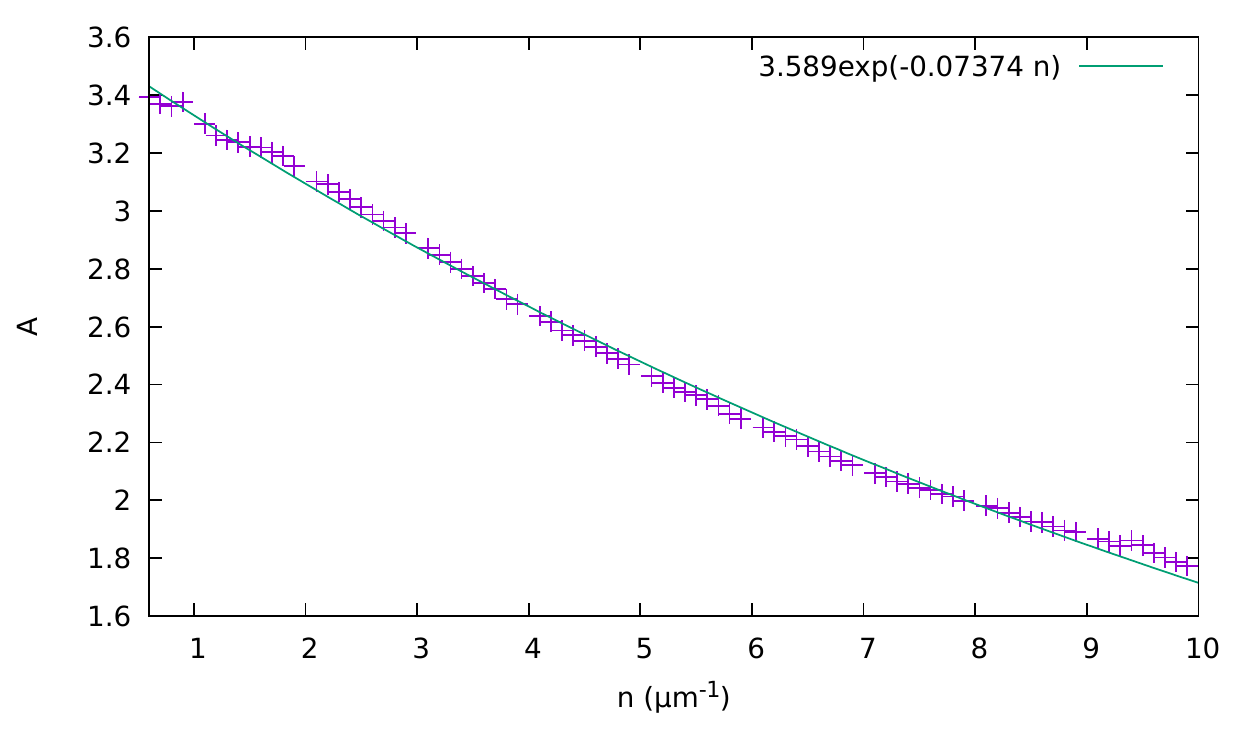}
  \caption{Behavior of $A$ obtained by fitting the variational
    $\gamma$ to an expression of the form~(\ref{eq:geff-def}) for a
    given density $n$. The amplitude $A$ is seen to decrease
    approximately exponentially with $n$, reaching a value of order
    $3.6$ at low density.  }
  \label{fig:A_vs_n}
\end{figure}

In Ref.~\onlinecite{tang2018}, the dipolar interaction is approximated by
replacing the full interaction~(\ref{V1d_dd1}) by
$g^{\mathrm{total}}_{1D}\delta(x)$ where:
\begin{eqnarray}
\nonumber
  \label{eq:geff-def}
  g^{\mathrm{total}}_{1D}=g_{VdW}+V(\theta) l_\perp \left(A-\frac 8 3\right) \delta(x)
  \\
\nonumber
  A=\int_{-\sqrt{2\pi}}^{\sqrt{2\pi}} du V_{DDI}^{1D} (u),
\end{eqnarray}
the integration bounds being chosen so that $A=3.6$ is 90\% of the
exact integral $\hat{V}_{DDI}(k=0)/l_\perp$. Analogously to our
variational approach the dipolar
interaction
is replaced by a simpler
contact interaction.  However, the criteria used to define the
contact interaction are markedly
different. In Ref.~\onlinecite{tang2018}, the potential is
replaced by a $\delta$-function potential
having almost the same Fourier transform at
$k=0$. Such approximation is expected to be valid when the
interparticle distance is large compared with $l_\perp$ , \emph{i.e.}
when $n l_\perp \ll 1$. In our variational approach, no assumption
is made on the two-particle scattering problem nor interparticle
distance, since the effective contact interaction is determined by the
minimization of the
energy per particle. At low density, $n l_\perp \ll 1$, it should be
possible to make contact with the approximation of
Ref..~\onlinecite{tang2018} and we should
have
\begin{eqnarray}
  \label{eq:limit-expected}
  \bar{g} \to A_{var} V(\theta) l_\perp,
\end{eqnarray}
where $A_{var}$ is a dimensionless constant. The values of
$A_{var}$ we find (see Fig.~\ref{fig:gamma_var}) are in reasonable agreement with
the approximations used in
Ref.~\onlinecite{tang2018}. However, the coefficient $A_{var}$ in
Eq.~(\ref{eq:limit-expected}) that gives the best fit to the variational
value of $\gamma$ decreases slightly with $n$. In
Fig.~\ref{fig:A_vs_n}, we show the dependence of $\tilde{A}$ obtained by
fitting the $\theta$ dependence of the optimal $\gamma$ with an
expression of the form~(\ref{eq:limit-expected}). We find that the
dependence can be described by an exponential, and that at low
density the result of Ref.~\onlinecite{tang2018} holds.

The minimum of the trial energy per particle,
Eq.~(\ref{eq:evar_k}), in units of $\frac{\hbar^2 n^2}{2m}$ is  $\epsilon_{var}(n)$, and can be expressed as
\begin{equation}
\epsilon_{var}(n)=\epsilon_{LL}[\gamma_0(n)]+\Delta \epsilon(n)
\label{eq:eperpar_var}
\end{equation}
where $\epsilon_{LL}(\gamma_0(n))$ is the dimensionless Lieb-Liniger
ground-state energy defined in Eq.~(\ref{eq:e0-lieb-liniger}))  at
dimensionless parameter $\gamma_0(n)$ (see Eq.~(\ref{eq:eff_gamma}))
while the contribution of the non-integrable soft-dipolar interaction
is encapsulated in the $\Delta \epsilon(n)$ correction.

\begin{figure}[h]
  \centering
  \includegraphics[width=9cm]{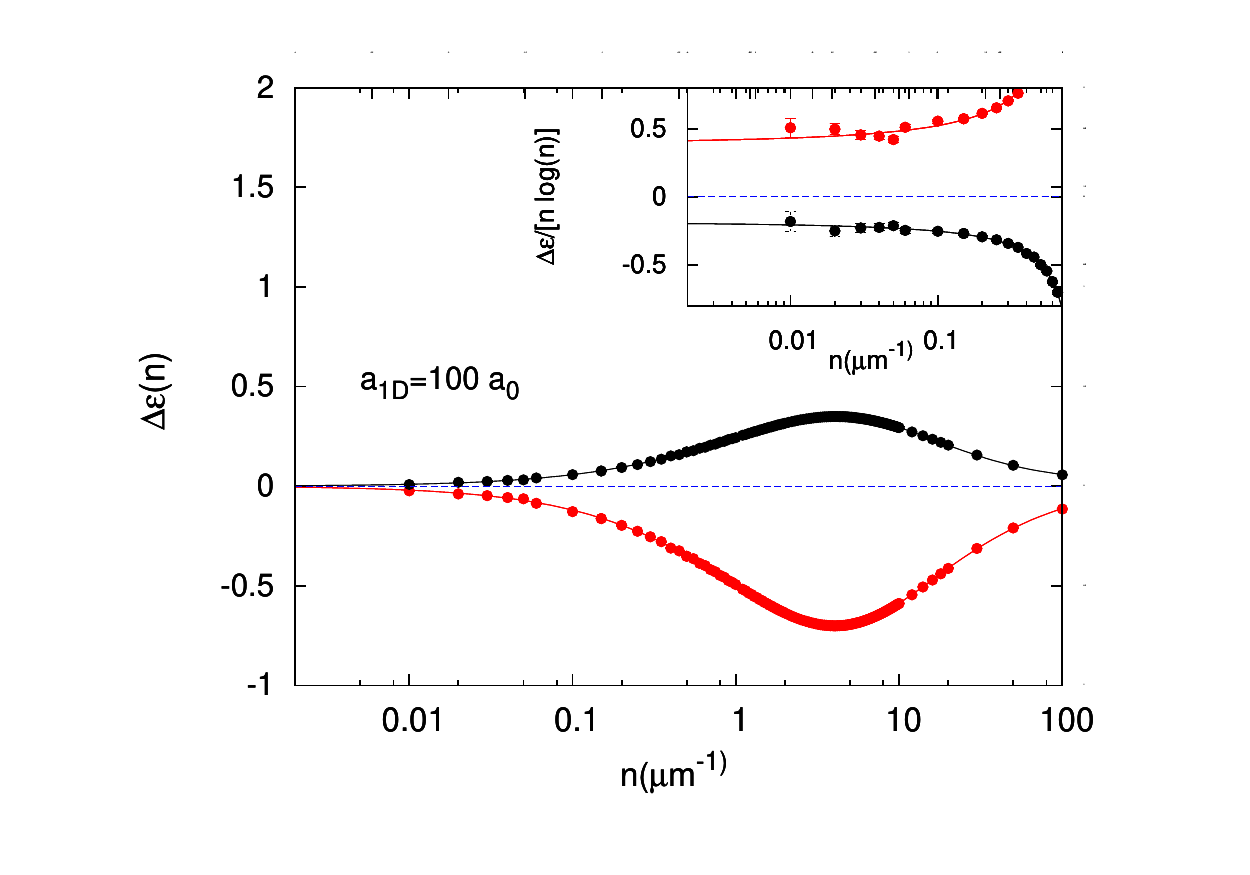}
  \caption{(Color online) Plot of $\Delta \epsilon(n)$ from Eq.~\ref{eq:eperpar_var} as a
    function of atom density for $a_{1D}=100 a_0$ for $\theta=0$ ( red solid dots)
    and $\theta=\pi/2 $ (black solid dots).
    The red and black solid lines are fitting curves of the form:
    $f(n)=(a n \log{n}+b n^{5/4}+c n^2)/(1+d n + e^{2+g})$. When $\theta=0$ we
    get $a=0.394(3),b=-0.551(6),c=-0.009(2),d=0.13(1),e=0.00039(8),g=1.14(1)$; when
    $\theta=\pi/2$ we get $a=-0.1853(2),b=0.287(1),c=-0.00136(4),d=0.168(5),e=-0.0005(1)
    ,g=0.73(3)$. In the inset we shown the low density behavior of
    $\Delta \epsilon(n)/(n \log{n})$  with the same color code.}
  \label{fig:e_log}
\end{figure}

As discussed before, at large density the $\gamma \rightarrow \gamma_0$ and
$\epsilon_{var}(n)\rightarrow \epsilon_{LL}(\gamma_0(n))$. At small density, whenever the optimal $\gamma$ is so large
that we can approximate the static structure factor as the one of the non-interacting fermionic gas,
the correction due the soft dipolar interaction $\Delta \epsilon(n) \propto n \log(n) $
(see Appendix ~\ref{app:high_and_low} for a detailed discussion).
This behavior can be seen in Fig.~\ref{fig:e_log} where the $\Delta \epsilon(n)$ corrections are shown
as a function of density for $a_{1D}=100 a_0$. At large density as well as for small density these corrections
go to zero, 
in the inset we show $\Delta \epsilon(n)(n \log(n))$ that for extremely low density goes towards a constant value.

\subsection{$\theta > \theta_c$: repulsive soft dipolar interaction}

\begin{figure}[h]
  \centering
  \includegraphics[width=9.5cm]{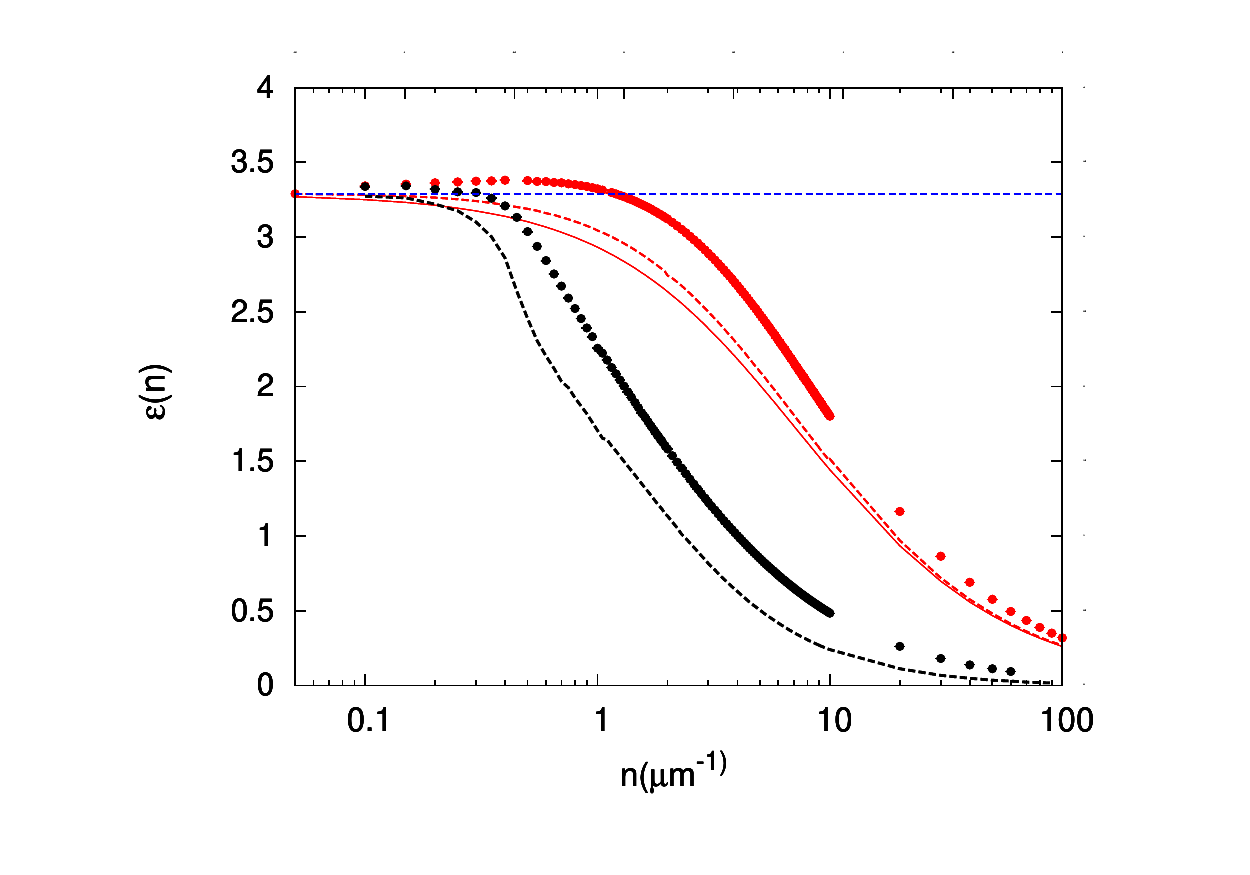}
  \caption{(Color online) Plot of energy per particle $\epsilon(n)$
  as a function of atom density, at
  $\theta=\pi/2$ corresponding to the largest repulsive soft dipolar interaction, for
  two scattering length $a_{1D}/a_0= 1000 $ and $10000 $;
  red and black solid dots respectively. With the same color code the
  $\epsilon_{LL}[\gamma_0(n)]$ (solid line) and $\epsilon_{LL}[\gamma(n)]$ (dashed line)
  are shown. $\epsilon_{LL}[\gamma_0(n)]$ for $a_{1D}/a_0=10000 $ is not shown since
  for the parameter we have chosen $\gamma_0(n) < 0$. The dashed blue
  line represents the low density limit $\pi^2/3$. }
  \label{fig:e_rep}
\end{figure}

As already observed, the overall effect of the repulsive soft dipolar
interaction is to make the optimum $\gamma(n) >\gamma_0$. In the large density limit,
$n l_\perp \gg 1$,
$\gamma \to \gamma_0$ and according to
Eq.~(\ref{eq:eff_gamma}), we can find $\gamma_0<0$ if
\begin{eqnarray}
  \label{eq:stability-repulsive}
  a_d>\frac{3 l_\perp^2}{4|a_{1D}|(1-3\cos^2\theta)}.
\end{eqnarray}
If that condition is satisfied, the
dipolar gas is unstable at high density. Otherwise, the gas is stable
for all densities.
In Fig.~\ref{fig:e_rep} we show $\epsilon_{var}(n)$, $\epsilon_{LL}[\gamma_0(n)]$
together with $\epsilon_{LL}[\gamma(n)]$, for various scattering length $a_{1D}$
as a function of the density.
At very low density the Tonks-Girardeau limit $\epsilon(n)=\pi^2/3$ is
recovered, while in the very high density limit the weakly
interacting regime is recovered. The quantities $ \epsilon_{LL}[\gamma_0(n)]$ and $ \epsilon_{LL}[\gamma(n)]$,
solid and dashed lines in Fig.~\ref{fig:e_rep} respectively,
can be seen as successive approximations to the variational energies (solid dots).
In Fig.~\ref{fig:e_rep} we have also shown the case ($a_{1D}/a_0=10000$, black data) where,
for the parameters chosen to make the calculation, $\gamma_0(n) < 0$
so the condition~(\ref{eq:stability-repulsive}) is met.
In the minimization procedure, for the densities considered, we always get
a positive optimal $\gamma$.

\subsection{$\theta < \theta_c$: attractive soft dipolar interaction}
When $V(\theta)$ is negative, $\gamma_0$ is enhanced by the contact contribution
and the effect of the soft dipolar interaction is to lessen this repulsion.
For small scattering lengths, the situation is similar to the one described for the
repulsive case but with a negative correction with respect to
$\epsilon_{LL}[\gamma_0(n)]$ in Eq.~(\ref{eq:eperpar_var}).

\begin{figure}[h]
  \centering
  \includegraphics[width=9cm]{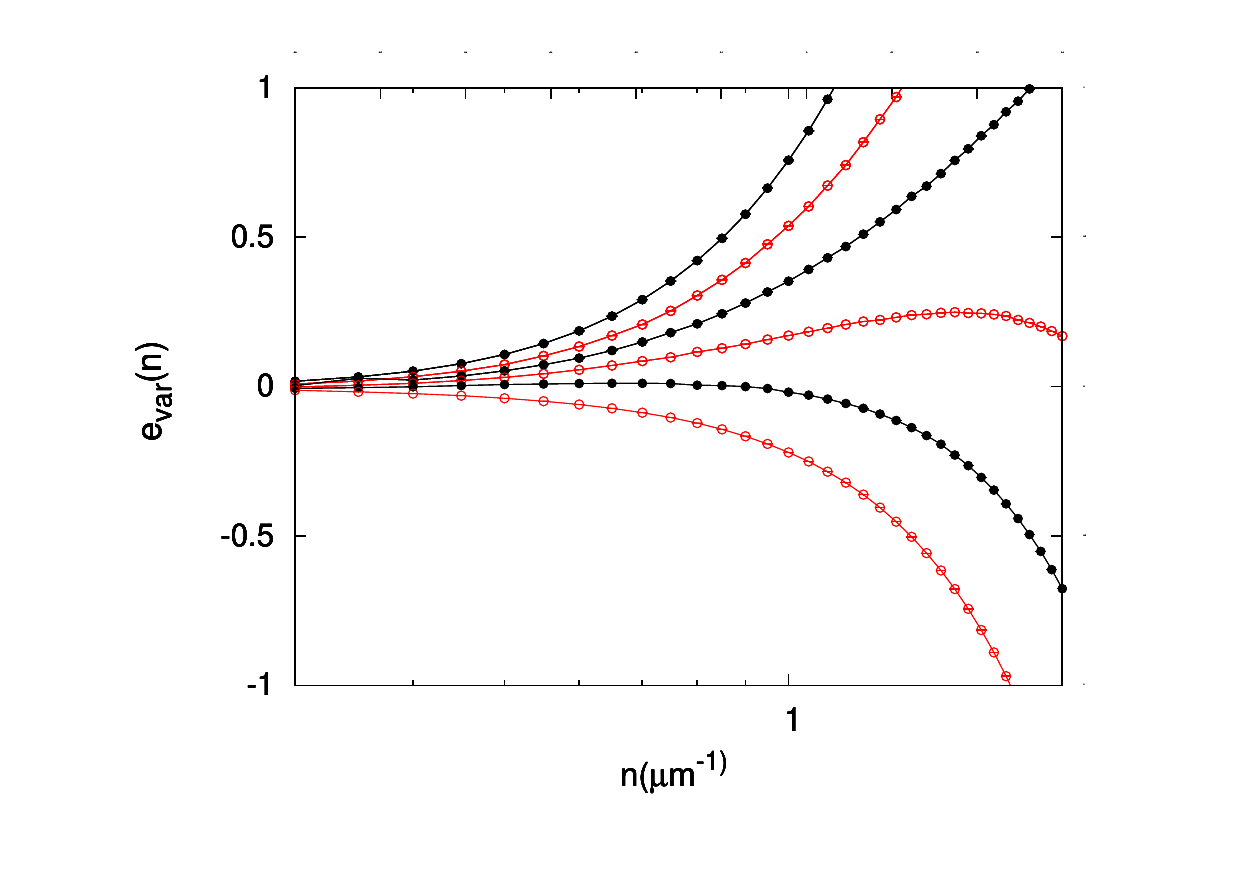}
  \caption{(Color online) Plot of energy per unit length as a function of atom density
   for selected scattering lengths. Starting from the top
   $a_{1D}/a_0=5000,6000,7000,8000,9000$ and $10000$ for fixed angle $\theta=0$. }
  \label{fig:e_att}
\end{figure}

However for large scattering lengths at small angles, when the effect of the
soft dipolar interaction is larger the system can become unstable.
In Fig.~\ref{fig:e_att} we show the variational ground state energy
per unit length $e_{var}(n)$ for different increasing scattering lengths.
When $a_{1D}/a_0> 8000$ the energy per unit length is convex at low density, but
presents a concavity at higher density. In such case, the compressibility becomes
negative, indicating an instability towards collapse at $n=n_c$ where the second derivative
of energy per unit length as a function of density is vanishing.

\subsection{Tomonaga-Luttinger parameters }
\label{sec:link2Lutt}

Having obtained the ground state energy per unit length, with
Eqs.~(\ref{eq:galilean}-\ref{eq:compress})
we can calculate the Tomonaga-Luttinger exponent $K$ as well as
the velocity of excitations $u$  that enter the bosonized Hamiltonian
Eq.~(\ref{eq:bosonized}).
\begin{figure}[h]
  \centering
  \includegraphics[width=9cm]{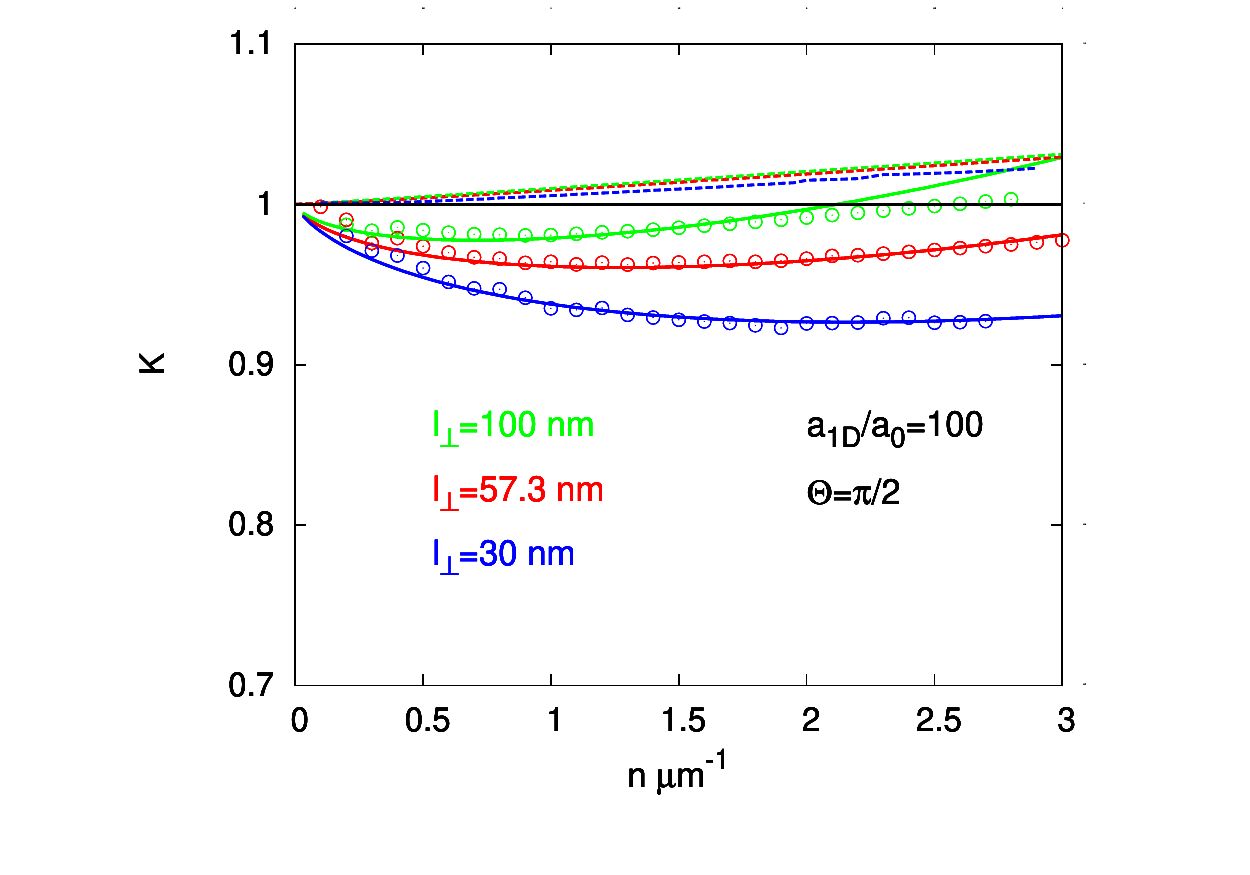}
  \caption{(Color online) Plot of the
    Tomonaga-Luttinger exponent $K$ deduced from the variational
    ground state energy in case of maximally repulsive dipolar interaction at $a_{1D}/a_0=100$
    for three different values of confinement: $l_\perp=100,57.3 nm $ and $30 nm$
    (blue, red and green data respectively).
    Open dots are $K$ obtained from $e_{var}(n)$ by numerical differentiation,
    solid curves are $K$ obtained by differentiating  the fitted
    expression  $e(n)=n^3(\pi^2/3+a n \log{n}+b n^c)$.
    The dashed curves are the Tomonaga-Luttinger exponents of the
    Lieb-Liniger gas computed
    at the optimal $\gamma$.}
  \label{fig:kll_a1d_20_lp}
\end{figure}
At very low density the Tomonaga-Luttinger exponent has a logarithmic correction
that qualitatively can be understood as follows (see also App.~\ref{app:high_and_low}).
In the limit of low density, one can approximate
  \begin{equation}
    \label{eq:hcb-gs}
    e_{GS}\simeq\frac{\pi^2 n^3}{6m} + \frac
  {n^2}{2} \hat{v}(k=0) + \int_0^{2\pi n} \frac{dk}{2\pi} \hat{v}(k)
  \left(\frac{k}{2\pi} -n \right),
\end{equation}
where we have used Eq.~(\ref{eq:hcb-sofk}), giving
\begin{equation}\label{TLexp-approx}
  K\simeq\frac 1 {\sqrt{1+\frac{m}{\pi^2 n} [\hat{v}(0)-\hat{v}(2\pi n)]} }.
\end{equation}

In the case of dipolar forces\cite{Reimann}, $v(x)$ behaves for long distance distance as
$\sim |x|^{-3}$, so $\hat{v}(0)-\hat{v}(2\pi n) \sim n^2 |\ln n|+ O(n^2)$, leading to $K-1 \sim n |\ln n|$.
This low density behavior can be traced in $K$ when $\gamma$
minimizing the variational energy is sufficiently large to
satisfy the above conditions. See for example
Fig.~\ref{fig:kll_a1d_20_lp},  where  $K(n)$ obtained by fitting the
low density behavior of the
$e_{var}(n)$ including a logarithmic correction (solid line) with an expression  $e(n) = n ( \pi^2/3 + a n \log{n} + b n^c)$,
is in agreement with the values obtained by numerical differentiation (open dots).

When the dipolar interaction is repulsive, the Tomonaga-Luttinger
exponent is lower than the exponent of a Lieb-Liniger gas having
either $\gamma=\gamma_0$  or $\gamma$ minimizing the variational energy
(\ref{eq:evar_k}).
This shows that the contribution from
non-integrable soft dipolar interaction in~(\ref{eq:decomposition}) is not
negligible, and that it is not correctly approximated by replacing the
non-contact interaction in~(\ref{eq:decomposition})  by an effective
contact interaction only.

Moreover, such approximations always lead to a Tomonaga-Luttinger exponent larger than
unity,\cite{cherny2008} and as we will see below, the dipolar gas can
have a Tomonaga-Luttinger exponent less than unity.
Reducing $l_\perp$  with fixed $r$, $V_{DDI}(r/l_\perp) \to (l_\perp/r)^3$ while $V(\theta)$ increases as
$l_\perp^{-3}$.  As the contribution to the ground-state energy is
enhanced, the Tomonaga-Luttinger exponent
is progressively reduced and for small density it can be less than
unity like in the strictly one-dimensional dipolar gas
dipolar gas.\cite{citro06_dipolar1d,depalo08_dipolar,citro07_luttinger_hydrodynamics,Roscilde2010}.
In Fig.~\ref{fig:kll_a1d_20_lp} we show $K(n)$ for a fixed $a_{1D}/a_0=100$ at $\theta=\pi/2$, varying the confinement:
namely $l_\perp=100, 57.3$ and $30$$nm$.
This is a clear indication that approximating the
Tomonaga-Luttinger exponent of the dipolar gas with the exponent of a
Lieb-Liniger gas can lead to results incorrect not just quantitatively
but also qualitatively. Indeed, finding a Tomonaga-Luttinger
exponent $1/2<K<1$ yields\cite{giamarchi_book_1d} $S(k \simeq 2\pi n) \simeq S(2\pi n)
+ C |k -2\pi n|^{2K-1} + o(|k -2\pi n|^{2K-1})$, giving a cusp
in $S(k)$ near $k=2\pi n$, whereas such cusp is absent with
$K>1$. However, the dynamical superfluid susceptibility remains
divergent as long as $K>1/4$.
\begin{figure}[h]
  \centering
  \includegraphics[width=9cm]{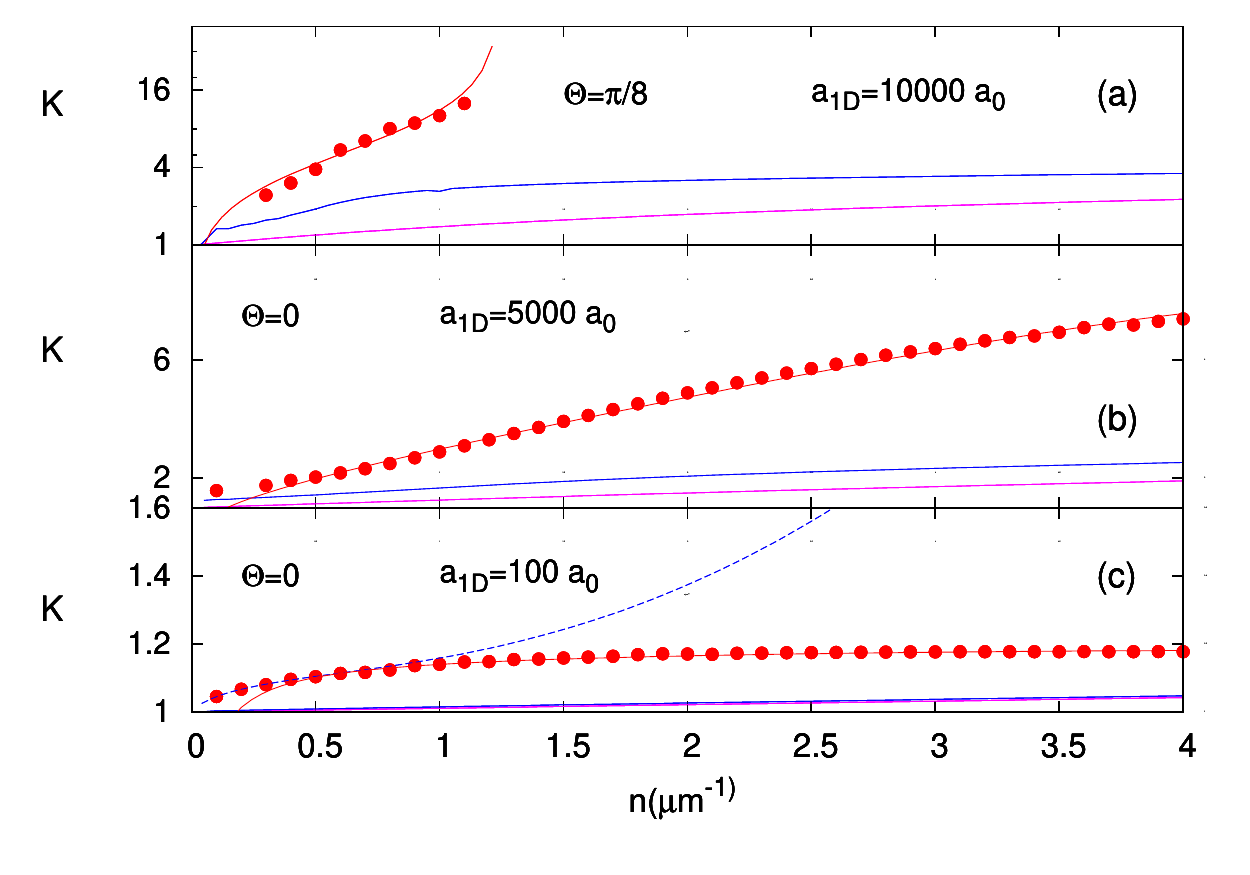}
  \caption{(Color online) Plot of the
    Tomonaga-Luttinger exponent $K$ derived from variational
    ground state energy in case of attractive dipolar interaction:
    $\theta < \theta_c$. Red solid dots represent $K$ obtained from $e_{var}(n)$
    by  numerical differentiation, solid red curve $K$
    obtained using a six degrees polynomial fit for the variational
    ground state energy and analytic differentiation.
    Blue and magenta solid curves are
    the Tomonaga-Luttinger exponents of a Lieb-Liniger gas computed
    respectively at the optimal $\gamma$ and at $\gamma_0$. In panel (a) the
    results are for $a_{1D}=10000 a_0$ and $\theta=\pi/8$ while in both panels
    (b) and (c)  $\theta=0$ with $a_{1D}=5000 a_0$ and
    $a_{1D}=100 a_0$ respectively. The blue dashed curve in panel (c) is
    a fit of the exponent obtained by numerical differentiation of
    $e_{var}(n)$ to an expression including logarithmic correction. }
  \label{fig:kll_zero}
\end{figure}
\begin{figure}[h]
  \centering
  \includegraphics[width=9cm]{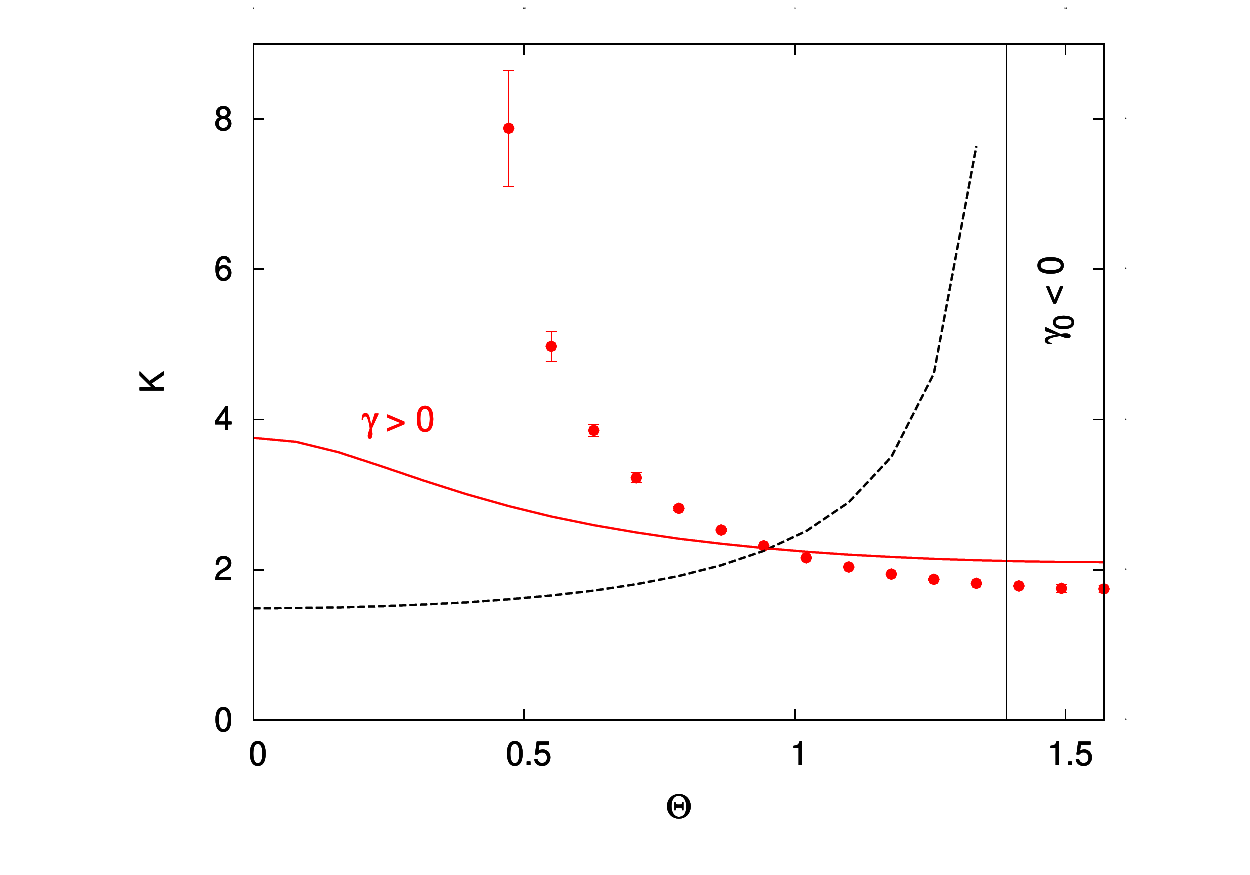}
  \caption{(Color online) Plot of the Tomonaga-Luttinger exponent $K$
    deduced  from the variational
    ground state energy as a function of $\theta$ at fixed density $n=1.5 \mu m^{-1}$,
    and scattering length $a_{1D}/a_0=10000$.
    A region of instability is found for $\theta_c < \pi/8$.
    The solid red points represent  $K$ obtained from  $e_{var}(n)$ by numerical differentiation,
     the solid red curve represents the Tomonaga-Luttinger exponents of the Lieb-Liniger gas computed
    at the optimal $\gamma$. The black dashed curve is $K$ in the
    Lieb-Liniger gas at $\gamma_0$.}
  \label{fig:kll_theta_a1d_10000_n15}
\end{figure}

In the attractive case, the Tomonaga-Luttinger exponent, which is related to the compressibility by Eq.~(\ref{eq:compress}), diverges
when the homogeneous ground state becomes unstable. In Fig.~\ref{fig:kll_zero} in panel (a), for $a_{1D}=10000 a_0$ we show the case in which
the system undergoes an instability with $K$ diverging at  $n_c \sim 1.2 \mu m^{-1}$.
Panels (b) and (c), for $a_{1D}/a_0=5000$ and $100$ respectively, show instead  the enhancement of $K$ due the
attractive interaction with respect
to the approximations using the Tomonaga-Luttinger exponent of
the Lieb-Liniger gas at the optimal $\gamma$ or at
$\gamma_0$. Moreover, when $a_{1D}/a_0=100$,
the optimal  $\gamma$ corresponds to the Tonks-Girardeau limit
(very low density), and the logarithmic correction to the exponent is visible (
see panel (c) where the blue dashed line is obtained by fitting the variational energy taking into account $n \log(n)$ term, while
a six degrees polynomial fit for the variational ground state energy gives results that do not match with the
values of $K(n)$ obtained with numerical differentiation).
In Fig.~\ref{fig:kll_theta_a1d_10000_n15} we follow the behavior of the Tomonaga-Luttinger exponent, at fixed
density $n=1.5 \mu m^{-1} $ and fixed $a_{1D}=10000 a_0$, as a function of the angle $\theta$. This a case where the original $\gamma_0$
is negative for large angles, while the optimal $\gamma$ is positive in the whole range of $\theta$.
However corrections beyond $e_{LL}(\gamma_{var})$ predicts an instability for $\theta < \pi/8$, as shown by the
diverging $K(n)$ corresponding to a change of sign in $\partial^2_{n}
e_{var}(n)$. While qualitatively, the Tomonaga-Luttinger exponent of a
Lieb-Liniger gas with the optimal $\gamma$ is a decreasing function of $\theta$ not showing any hint of
instability. Moreover, the exponent that would be obtained by
neglecting the non Lieb-Liniger part of Eq.~(\ref{eq:decomposition})
is an increasing function of $\theta$ and shows the instability for
$\theta>\frac {\pi}3$.

\section{Conclusion}
\label{sec:conclusions}
We have presented a variational approach to the ground state energy of
bosons in the continuum interacting by a two-body potential,
based on analytic expressions for
the ground state energy\cite{lang2017} and structure
factor\cite{cherny2008} of the Lieb-Liniger model.

Using this variational approach we have calculated the ground-state energy
bosonic atoms in transverse harmonic trapping with dipolar
interaction\cite{tang2018} treated within
the single mode approximation, as a function of density for several scattering
lengths,  confinement lengths and spanning the angle of dipoles.
From the ground-state energy we have estimated the Tomonaga-Luttinger
exponent as a function of density
and interaction: when dipolar interactions are attractive and density is sufficiently
high an instability of the Tomonaga-Luttinger liquid is predicted.
Knowledge of the dependence of the variational ground state energy on
the density will permit to consider
the effect of longitudinal harmonic trapping, in particular to compute the frequencies of the breathing
modes.\cite{menotti_stringari,fuchs2003,fuchs2004,oldziejewski_strongly_2019} This will be
the object of a future work. The variational method can be
applied to other systems of interest such as atoms interacting via
shoulder potentials\cite{rydberg_atoms} or power law
interactions.\cite{douglas2015}
The variational approach of the present paper could be extended in different directions. Since an exact form factor
representation for the structure factor\cite{caux_density} of the
Lieb-Liniger model is available, the
the Tomonaga-Luttinger exponent and critical density could be
calculated more accurately, albeit with greater
computational cost with respect to the semi-analytical approach presented here.  Second,
the variational principle used here can be
generalized to positive temperature\cite{feynman_statmech} and the
free energy of integrable models can be obtained from the
Thermodynamic Bethe Ansatz.\cite{yang1969} Using form factor expansion
techniques, the static structure factor of the Lieb-Liniger gas
has been calculated for
positive temperatures,\cite{panfil2014} thus the present variational
approach could also be generalized to free energy calculations for
positive temperatures.

\begin{acknowledgments}
  We thank B. Lev for discussions that inspired this
  project, \eo{and Z. Ristivojevic, L. Sanchez-Palencia, R. Oldziejewski for
  comments on the manuscript}. E. O. thanks SISSA and Universit\`a di Trieste for
  hospitality.
\end{acknowledgments}

\appendix
\section{Contributions to the variational energy}\label{app:pcf}

Writing explicitly the average by introducing $\rho(x)=\sum_{i=1}^N \delta (x-x_i)$ and $n=N/L$, being $N$ the number of particles and $L$ the system length, one has
\begin{equation}\label{var-pot}
\langle\psi_0(g)|\mathcal{V}|\psi_0(g)\rangle=\frac{Ln^2}{2} \int dx
v(x) g(x)
\end{equation}
where  $g(x)$ is the pair correlation function $n^2g(x)=\langle \psi_0(g)| \rho(x) \rho(0)|\psi_0(g)\rangle-\frac N L \delta(x)$.
In the integral~(\ref{var-pot}), one can introduce the static
  structure factor $S(k)$  with
\begin{equation}\label{strucfac}
g(x)=1+ \int_{-\infty}^{\infty} \frac{dk}{2\pi n} e^{ikx}\lbrack
S(k)-1\rbrack .
\end{equation}
where the relation is derived by the definition of the static structure factor:

\begin{eqnarray}
  \label{eq:strucfac-def}
  S(q)=\frac 1 n \int_{-\infty}^{+\infty} dx e^{-iq x} \left[\langle \rho(x)
  \rho(0)\rangle - n^2 \right],
\end{eqnarray}
so that the following relation between the static structure factor and the pair correlation function holds:
\begin{eqnarray}
S(q) &=& 1+n\int_{-\infty}^{+\infty} \left[ g(x)-1\right]e^{- i q x} dx 
\end{eqnarray}
and hence:
\begin{equation}
\!\!\frac{\langle \psi_0(g)|\mathcal{V}|\psi_0(g)\rangle }{L}\!=\!\frac{n}{2}\! \int_0^{+\infty}
\!\frac{dq}{\pi}  \hat{v}(k) [S(q)-1]+ \frac{n^2} 2 \hat{v}(q=0),
\label{pot_sofk}
\end{equation}
where we have used the parity $S(q)=S(-q)$ and
$\hat{v}(q)=\hat{v}(-q)$.

The other contribution to the variational energy is evaluated using the
Hellmann-Feynman theorem. One can show that in the
Lieb-Liniger model,
\begin{eqnarray}
  \frac{\langle \psi_0(g)|\mathcal{U} |\psi_0(g)\rangle} L = \frac {n^2} 2 \epsilon'(\gamma),
\end{eqnarray}
where
\begin{eqnarray}
\epsilon'(\gamma)=g(0).
\end{eqnarray}

\section{Ground state energy and static structure factor of the
  Lieb-Liniger gas }
\label{app:Lieb_liniger_results}

We can express the ground state energy $e_0(g)$ as a function of the
dimensionless parameter
$ \gamma=\frac{m g}{\hbar^2n} $ so that $e_0(g)=\frac{\hbar^2 n^3}{2 m} \epsilon_{LL}(\gamma)$.

The function $\epsilon$  is given by the solution of integral
equations\cite{lieb_bosons_1D}
\begin{eqnarray}
  \label{eq:BAE}
  2\pi \rho(\mu)&=&1+2c \int_{-q_0}^{q_0} d\mu'
  \frac{\rho(\mu')}{c^2 + (\mu-\mu')^2}, \\
\label{eq:BAdensity}
 n&=& \int_{-q_0}^{q_0} d\mu \rho(\mu),
  \\
  n^3 \epsilon_{LL}(\gamma)&=&
  \int_{-q_0}^{q_0} d\mu \mu^2 \rho(\mu),
\label{eq:BAenergy}
\end{eqnarray}
where $c=\frac{mg}{\hbar^2}=n\gamma$.

In our manuscript, instead of solving the integral
equations~(\ref{eq:BAE})--(\ref{eq:BAdensity}) we will use the approximate analytical
expression  for the  dimensionless
energy $\epsilon_{LL}(\gamma)$ suggested in Ref.~\onlinecite{lang2017}.

At small $\gamma$ where the Lieb-Liniger energy $\epsilon_{LL}$ is approximated by \cite{lang2017}
\begin{eqnarray}
\label{eLL_small}
\nonumber
&&\epsilon_{LL}(\gamma)=\gamma-\frac{4}{3\pi} \gamma^{3/2}+\left[ \frac{1}{6}-\frac{1}{\pi^2}\right]- 0.002005\gamma^{5/2}\\
&+&0.000419 \gamma^3 -0.000284 \gamma^{7/2}+0.000031 \gamma^4.
\end{eqnarray}
while from strong to intermediate coupling regime\cite{lang2017}
\begin{eqnarray}
\label{eLL_large}
\frac{\epsilon_{LL}(\gamma)}{\epsilon^{TG}}&=& \frac{\gamma^2}{(2+\gamma)^2}+
\sum^{\infty}_{n=1}\frac{\pi^{2n} \gamma^2}{(2+\gamma)^{3 n+2}} {\cal L}_n \\
\nonumber
{\cal L}_1 &=& \frac{32}{15} \\
\nonumber
{\cal L}_2 &=&- \frac{96}{35} \gamma +\frac{848}{315}\\
\nonumber
{\cal L}_3 &=& \frac{512}{105} \gamma^2 -\frac{4352}{525} \gamma +\frac{13184}{4725}\\
\nonumber
{\cal L}_4 &=&-\frac{1024}{99} \gamma^3 + \cdots
\end{eqnarray}
with $\epsilon^{TG}=\pi^2/3$. \eo{The expression (\ref{eLL_small}) was
  obtained\cite{lang2017} by fitting the ground state energy to a polynomial
  expression in $\sqrt{\gamma}$ for $\gamma<15$. An exact expansion has been
  conjectured\cite{ristivojevic2019,marino2019}. We have checked that
  in the range $0<\gamma<8$ the relative difference between the two
  expressions 
  was under $2\times 10^{-3}$, while the relative difference between
  the derivatives was under $10^{-2}$. Using the
  expansion\cite{ristivojevic2019,marino2019}
  instead of Eq.~(\ref{eLL_small})in the variational
  calculation of the ground state energy yields relative differences
  under $5\times10^{-3}$. Concerning the
  expression (\ref{eLL_large}), we note that an alternative asymptotic
expansion\cite{ristivojevic2014,lang2019} in powers of $1/\gamma$ also
applies for $\gamma\gg 1$.
However, the expression~(\ref{eLL_large}) is more convenient\cite{lang2017}
to match with (\ref{eLL_small}) in the intermediate region of $\gamma
\sim 1$.}  

Using the phenomenological suggestion given in Ref.~\onlinecite{cherny2008} for structure factor $S(k,\omega)$
it is possible to have an approximate estimate of the static structure factor $S(k)$ in terms of ratio
between Gauss hypergeometric functions\cite{abramowitz_math_functions}
\begin{widetext}
\begin{equation}
  \label{eq:sofk-hyper}
  S(k)=\frac{k^2}{2 m \omega_-(k)}
  \frac{{}_2F_1\left(1+\frac{\sqrt{K}}{1+\sqrt{K}}+\mu_-(k)+\mu_+(k),1+\mu_-(k);2+\mu_-(k)-\mu_+(k),1-\left(\frac{\omega_-(k)}{\omega_+(k)}\right)^2\right)}{{}_2F_1\left(1+\frac{2\sqrt{K}}{1+\sqrt{K}}+\mu_-(k)+\mu_+(k),1+\mu_-(k);2+\mu_-(k)-\mu_+(k),1-\left(\frac{\omega_-(k)}{\omega_+(k)}\right)^2\right)}
\end{equation}
\end{widetext}
where $K=4\pi^2\rho(\pm q_0)$ is the Tomonaga-Luttinger exponent, $\omega_+(k)$ is the
dispersion of the Type-I Lieb excitations\cite{lieb_excit}, $\omega_-(k)$ the
dispersion of the Type-II Lieb excitations\cite{lieb_excit} for
$k<2\pi n$ and $\omega_+(k-2\pi n)$ otherwise, and $\mu_\pm(k)$ are
the exponents of the threshold singularity respectively at
$\omega_\pm(k)$ and can be calculated from the shift function.\cite{khodas2007,imambekov2009,cherny2008} The
expression~(\ref{eq:sofk-hyper}) reduces to  Eq.~(22) in
Ref.~\onlinecite{cherny2008} when the approximation $K\simeq 1$ is
made.

\begin{figure}
  \centering
  \includegraphics[width=9cm]{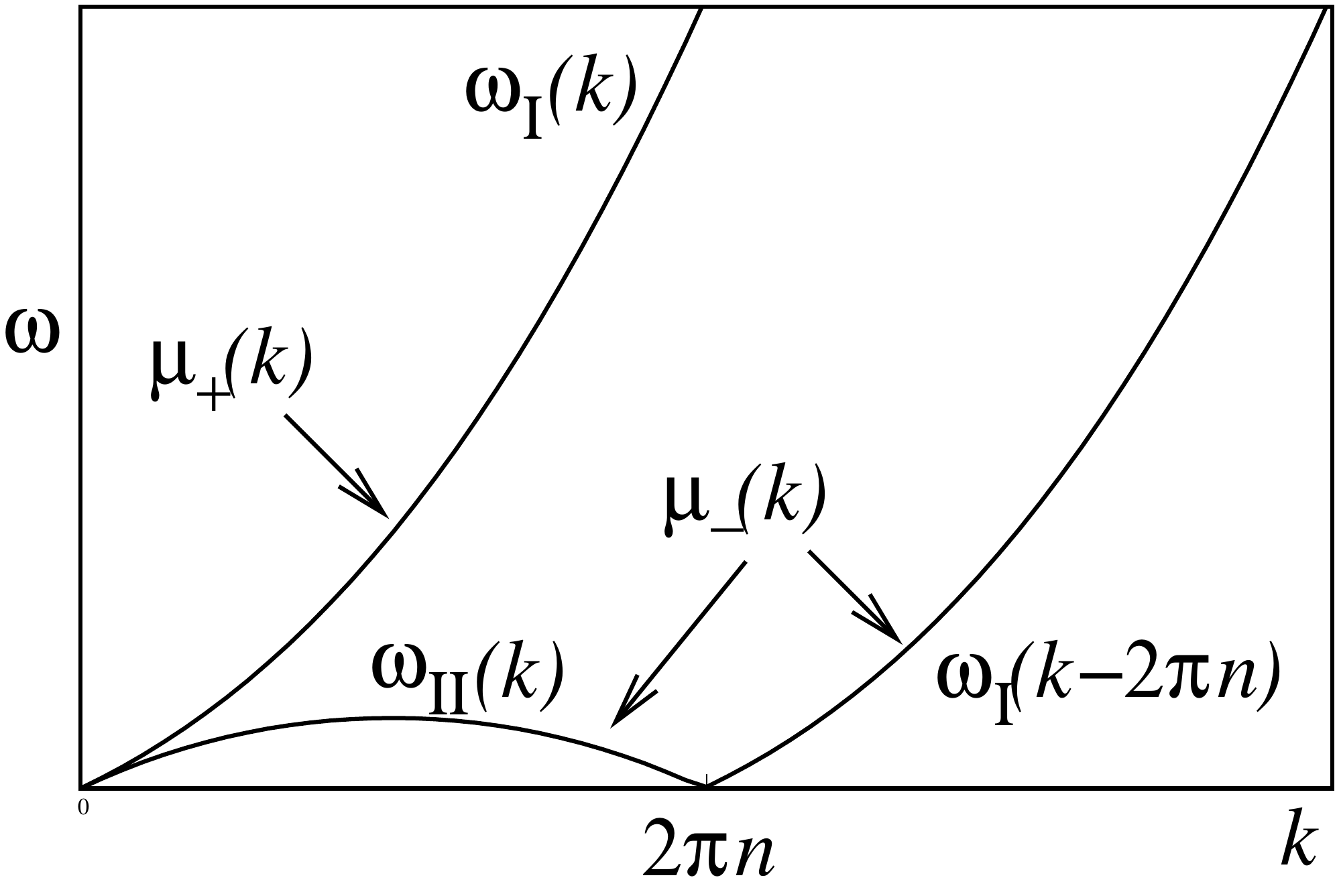}
  \caption{Dispersion of the Lieb-I and Lieb-II modes. The dynamic
    structure factor is non-vanishing for
    $\omega_{II}(k)<\omega<\omega_I(k)$ for $0<k<2\pi n$ and
    $\omega_I(k-2\pi n)<\omega<\omega_I(k)$ for $k>2\pi n$. $\mu_+(k)$
  is the threshold singularity exponent of the dynamical structure
  factor near the higher branch of
  excitations, while $\mu_-(k)$ is the exponent near the lower branch. }
  \label{fig:lieb-dispersion}
\end{figure}

To calculate $S(k)$ we have to consider\cite{cherny2008} the integral equation for the
shift function:
\begin{eqnarray}
  \label{eq:shift-function}
&&  F_B(\nu|\lambda)-\frac 1 {2\pi} \int_{-q_0}^{q_0}
  \frac{2c}{(\nu-\mu)^2+c^2} F_B(\mu|\lambda)\nonumber \\ && =\frac 1 2 + \frac 1 \pi
  \arctan\left(\frac{\nu-\lambda} c\right),
\end{eqnarray}
and
\begin{eqnarray}
&&  \omega_{p,h}(\lambda)=\pm \frac{ \hbar^2}{2m}\left[\lambda^2
  -2\int_{-q_0}^{q_0} \mu F_B(\mu|\lambda) d\mu \right]\\
&&  k_{p,h}(\lambda)=\pm \left[\lambda + 2 \int_{-q_0}^{q_0}
  \arctan\left(\frac{\lambda -\mu } c\right) \rho(\mu) d\mu\right]
\end{eqnarray}
Then, the dispersion of  Lieb modes is given by:
\begin{eqnarray}\label{eq:lieb-I}
  \omega_+(\lambda)&=&\omega_h(q_0) +\omega_p(q_0+\lambda) \nonumber \\
  k_+(\lambda)&=&k_h(q_0)+k_p(q_0+\lambda)
\end{eqnarray}
for type I, and
\begin{eqnarray}\label{eq:lieb-II}
  \omega_-(\lambda)&=&\omega_p(q_0) +\omega_h(q_0-\lambda) \nonumber \\
  k_-(\lambda)&=&k_p(q_0)+k_h(q_0-\lambda)
\end{eqnarray}
with $\lambda<q_0$ for type-II. When $\lambda>q_0$,
Eq.(\ref{eq:lieb-II}) reduces, up to a sign, to  the dispersion of the
Lieb-I mode
shifted of $2\pi n$.  For a given wavevector $k$, one must
first determine $\lambda_\pm(k)$ that solves $k_\pm(\lambda_\pm)=k$,
and calculate $\omega_{\pm}(\lambda_\pm(k))$ to obtain the
dispersion\cite{lieb_excit,cherny2008}. Once $\lambda_+$ is known, the
quantities
\begin{eqnarray}
  \label{eq:delta-def}
  \delta_{\pm}(\lambda_+(k))=2\pi F_B(\pm q_0,\lambda_+(k))
\end{eqnarray}
are found from the integral
equation~(\ref{eq:shift-function}), and the threshold exponent of the
Lieb-I mode\cite{khodas2007,imambekov2009} is
\begin{eqnarray}
  \label{eq:mu-plus-def}
  \mu_+(k)=1-\frac 1 2 \left(\frac 1 {\sqrt{K}} +
  \frac{\delta_+(\lambda_+)-\delta_-(\lambda_+)}{2\pi}\right)^2
  \nonumber \\ -\frac 1 2
  \left(\frac{\delta_+(\lambda_+)+\delta_-(\lambda_+)}{2\pi}\right)^2
\end{eqnarray}
Similarly, having found $\lambda_-$, one obtains $\delta_\pm(\lambda_-)$ by
replacing $\lambda_+$ with $\lambda_-$ in (\ref{eq:delta-def}).
Substituting $\lambda_+$ with $\lambda_-$ in
Eq.(\ref{eq:mu-plus-def}),
the threshold singularity  exponent $\mu_-(k)$
of the dynamical structure factor at the
lower edge is found.\cite{khodas2007,cherny2008,imambekov2009} For $k<2\pi n$,
the lower edge is given by the Lieb-II mode, while for $k>2\pi n$ it
is given by a replica of the Lieb-I mode shifted by $2\pi n$.

\begin{figure}[h]
\begin{center}
\includegraphics[height=65.mm]{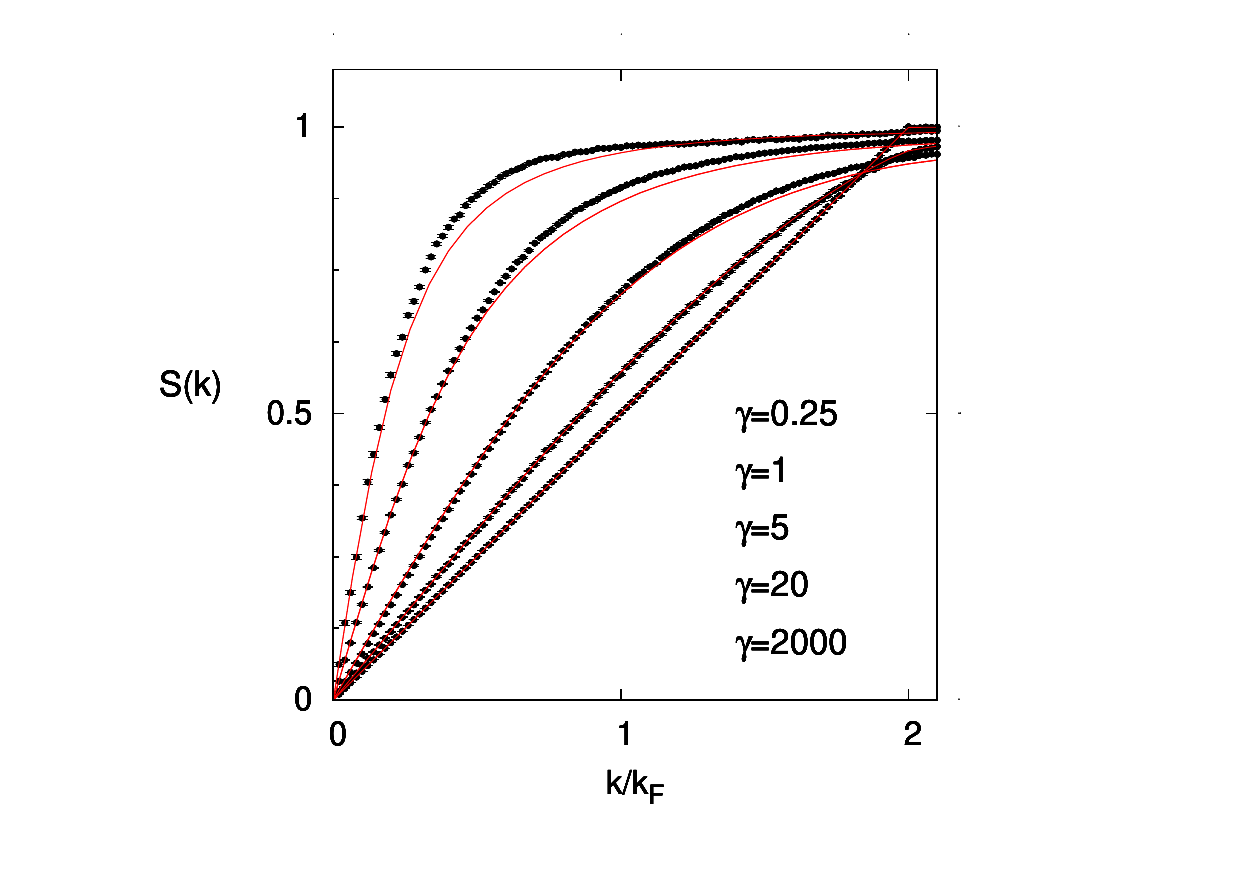}
\end{center}
\caption{Comparison of the structure factor $S(k)$ calculated from the
  Cherny-Brand Ansatz\cite{cherny2008} (red solid lines) with the
  structure factor obtained from Quantum Monte Carlo calculations
  (black solid dots).  $S(k)$ is increasing linearly for low $k$ with a
slope that is a decreasing function of $\gamma$. For large $k$, it
saturates to the value $1$. In the figure, $k_F=\pi n$.}
\label{fig:cfr_sq}
\end{figure}

In the limit of
$\gamma \to +\infty$, the exact expression of $S(k)$ simplifies to
\begin{equation}
  \label{eq:hcb-sofk}
  S(k)=\frac{|k|}{2\pi n} \theta(2\pi n -|k|) + \theta(|k|-2\pi n).
\end{equation}
In some selected cases, $S(k)$ from Quantum Monte Carlo simulations
can be  used
as benchmark for expression~(\ref{eq:sofk-hyper}) ; comparison between the static structure factor from simulations and from the approximated
Ansatz,for some selected cases, are shown in
Fig.~\ref{fig:cfr_sq}. The linear increase $S(k)=\frac{K |k|}{2\pi n}
+ o(k)$ for small $k$ predicted by
bosonization\cite{citro06_dipolar1d} is well reproduced by both
simulations and the Ansatz~(\ref{eq:sofk-hyper}). The agreement
between simulations and Ansatz is better for the case of $\gamma>1$.
We note that in Ref.~\onlinecite{cherny2008}, a satisfactory
comparison with the expression obtained for form factor summation was
already shown.

\section{minimization of the energy}\label{app:minimization}
\begin{figure}[h]
\begin{center}
\includegraphics[height=65.mm]{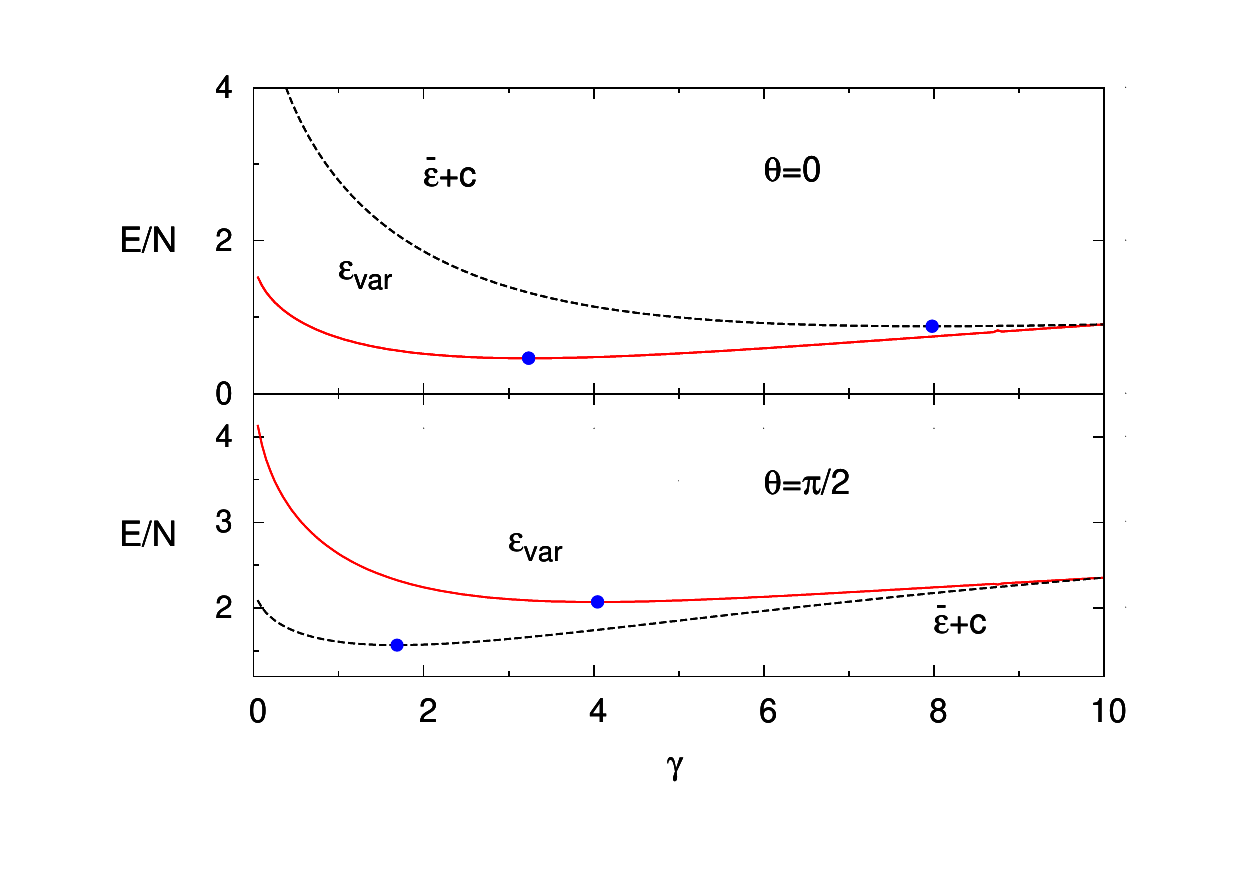}
\end{center}
\caption{$E/N$ as a function of $\gamma$ with ( red solid line) and without the soft dipolar interaction, $\bar{\epsilon}=
\epsilon(\gamma)-\bar{\gamma}\frac{\partial \epsilon(\gamma)}{\partial \gamma}$ where we have added a constant
(black dashed line). The solid blue points on the curves show the minimum of $E/N$.
In the upper panel we show results for $\theta=0$ when $V^{1D}_{DDI}$ is maximally attractive and
in the lower panel we show results for $\theta=\pi/2$ when $V^{1D}_{DDI}$ is maximally repulsive }
\label{fig:evar_n20_a1d_5000}
\end{figure}

In Fig.~\ref{fig:evar_n20_a1d_5000} we show
Eq.~(\ref{eq:evar_k}) with and without the dipolar interaction term, for $a_{1D}=5000 a_0$
and $n=2.0 \mu m^{-1}$, for $\theta=0$ when the interaction is maximally attractive
and for $\theta=\pi/2$ where when the interaction is maximally repulsive.
When the dipolar interaction is repulsive it enhances the total
repulsion and hence the optimal value of $\gamma$ its greater than
$\gamma_0$, whereas when it is
attractive it reduces the total repulsion so that the optimal $\gamma
< \gamma_0$.

\section{high and low density limits of the variational energy}
\label{app:high_and_low}
Analytic expressions of $\frac{E}{L}$ can be obtained using the
  Lieb-Liniger ground state energy derived in \cite{lang2017} and the
  expressions of the structure factor derived in \cite{cherny2008}
  for $\gamma\ll 1$ and for $\gamma \gg 1$. In the latter case, using
  (\ref{eq:hcb-gs}), we have
\begin{eqnarray}
&&\frac{E}{L}=\frac{\hbar^2 \pi^2 n^3}{6m} +\frac{\hbar^2 n^2 a_d}{m
  l_\perp^2} (1-3 \cos^2 \theta) +\frac{\hbar^2 a_d (1-3\cos^2
  \theta)}{\pi m l_\perp^4}  \nonumber \\ && \times \int_0^{2\pi n l_\perp} du \left(\frac u
  {2\pi} - n l_\perp\right) \left[1-\frac {u^2} 2 e^{\frac {u^2} 2}
    E_1\left(\frac{u^2}2\right) \right],
\end{eqnarray}
and using (\ref{eq:galilean})--~(\ref{eq:compress})
the Tomonaga-Luttinger exponent is given by
\begin{equation}
\label{eq:TLL-sinha-hcb}
\!\! K^{-2}\!\!= \!1 +\! n a_d (1-3\cos^2 \theta) e^{\frac{(2\pi n l_\perp)^2} 2}
  E_1\left(\! \frac{(2\pi n l_\perp)^2} 2 \right),
\end{equation}
so that in the limit $n l_\perp \to 0$, $K=1+n a_d (1-3\cos^2 \theta)
\ln (\pi \sqrt{2} e^{\gamma/2} n l_\perp) + O(n^3)$, the expected
behavior with interactions decaying as $1/|x|^3$ at long distance.
In the case of $\gamma \ll 1$, we can approximate\cite{cherny2008}
\begin{eqnarray}
  \label{eq:bogoliubov_Sk}
  S(k)\simeq \frac{|k|}{\sqrt{k^2 + 4\gamma n^2}},
\end{eqnarray}
leading to a variational energy
\begin{eqnarray}
  \label{eq:bogo-energy}
  \frac E L &\simeq& \frac{\hbar^2 n^3}{2m} \left[\epsilon (\gamma) -\bar{\gamma}
  \frac{\partial \epsilon}{\partial \gamma}\right]  + V(\theta) \left[ 2 n^2
  l_\perp - n \sqrt{\frac \pi 2} \right. \nonumber \\&& \left.+ \frac{n}{2\pi} \int_0^{+\infty}
  \frac{1-u e^u E_1(u)} {\sqrt{2 u + 4 \gamma n^2 l_\perp^2}} du
  \right].
\end{eqnarray}
Minimizing with respect to $\bar{\gamma}$, we obtain

\begin{equation}
  \label{eq:der-energy}
  \frac{\partial}{\partial \gamma}\left(\frac E L\right) \simeq -
  \frac{\hbar^2 n^3}{2m} \bar{\gamma}\epsilon''(\gamma) +
  \frac{n}{2\pi} \frac{dI}{d\gamma}(\gamma) =0,
\end{equation}
where we have defined
\begin{equation}
I(\gamma) = \int_0^{+\infty}
  \frac{1-u e^u E_1(u)} {\sqrt{2 u + 4 \gamma n^2 l_\perp^2}} du
\end{equation}

According to Eq. (\ref{eq:gamma-def}), for $n\to +\infty$, $\gamma n^2 \to
+\infty$, so we can replace the denominator in the integral $I(\gamma)$  with
$(4\gamma n^2 l_\perp^2)^{3/2}$. Using the expansion\cite{lang2017} of
$\epsilon(\gamma)$ valid for $\gamma \ll 1$, we end up with
\begin{eqnarray}
  \bar{\gamma} \simeq \sqrt{2 \pi} \frac{
  \hbar^2 a_d (1-3\cos^2 \theta)}{m g n^2 l_\perp^4},
\end{eqnarray}
so $\bar{g}=O(n^{-1})$ when $n \to +\infty$. In the high density
limit, $\gamma \to \gamma_0$, and the
leading order expansion is
\begin{eqnarray}
  g &=&g_{VdW} -  \frac{\hbar^2 a_d(1-3 \cos^2 \theta)}{2m
  l_\perp^2} \\ && \times \left[\frac 8 3 -\sqrt{\frac \pi 2} \frac{1}{n a_d
  \left(\frac{4l_\perp^2}{a_d |a_{1D}|} - \frac 8 3 (1-3\cos^2
  \theta)\right)}+\ldots \right]    \nonumber
\end{eqnarray}
This behavior is illustrated on Fig.~\ref{fig:gamma_n_a1d}.

\end{document}